%Paper: hep-th/9510114
%From: palla@ludens.elte.hu
%Date: Tue, 17 Oct 1995 15:09:00 +0100

\magnification1200
\font\BBig=cmr10 scaled\magstep2

\font\BBBBig=cmr10 scaled\magstep4
\font\small=cmr7

%%%%%%%%%%%%%%%%%%%%%%%%%%%%%%%%%%%%%%%%%%%%%%%%%%%%%%
%%%%%%%%%%%%%%%%%%%%% the title %%%%%%%%%%%%%%%%%%%%%%
%%%%%%%%%%%%%%%%%%%%%%%%%%%%%%%%%%%%%%%%%%%%%%%%%%%%%%

\def\title{
{\bf\BBBBig
\centerline{Spinors in non-relativistic}
\bigskip
\centerline{Chern-Simons electrodynamics}
}}%%%%% for the front page

\def\foot#1{
\footnote{($^{\the\foo}$)}{#1}\advance\foo by 1
} %%%%% foot(notes
\def\ccr{\cr\noalign{\medskip}}

%%%%%%%%%%%%%%%%%%%%%%%%%%%%%%%%%%%%%%%%%%%%%%%%%%%%%%
%%%%%%%%%%%%%%%%%%%% the author(s) %%%%%%%%%%%%%%%%%%%
%%%%%%%%%%%%%%%%%%%%%%%%%%%%%%%%%%%%%%%%%%%%%%%%%%%%%%

\def\authors{
\centerline{
{C.~DUVAL}\foot{
D\'epartement de Physique, Universit\'e de la M\'editerrann\'ee
and
Centre de Physique Th\'eorique, CNRS-Luminy, Case 907,
F--13288 MARSEILLE,
Cedex 09 (France).
E-mail: duval@cpt.univ-mrs.fr
\hfill\break
}
{P.~A.~HORV\'ATHY}\foot{
Laboratoire de Math\'ematiques et Applications,
Universit\'e de Tours, Parc de Grandmont,
F--37200 TOURS (France).
E-mail: horvathy@balzac.univ-tours.fr
\hfill\break
}
{L.~PALLA}\foot{
Institute for Theoretical Physics,
E\"otv\"os University, H--1088 BUDAPEST, Puskin u. 5-7 (Hungary).
E-mail: palla@ludens.elte.hu}
}
}

\def\runningauthors{Duval, Horv\'athy, Palla}

\def\runningtitle{Spinors in Chern-Simons \dots}

%%%%%%%%%%%%%%%%%%%%%%%%%%%%%%%%%%%%%%%%%%%%%%%%%%%%%%
%%%%%%%%%%%%%%%%%%%% the metrics %%%%%%%%%%%%%%%%%%%%%
%%%%%%%%%%%%%%%%%%%%%%%%%%%%%%%%%%%%%%%%%%%%%%%%%%%%%%

\voffset = 1cm %%%%% ver printing
\baselineskip = 20pt %%%%% line spacing

\headline ={
\ifnum\pageno=1\hfill
\else\ifodd\pageno\hfil\tenit\runningauthors\hfil\tenrm\folio
\else\tenrm\folio\hfil\tenit\runningtitle\hfil
\fi
\fi}

\nopagenumbers
%\pageno=1
\footline={\hfil} %%%%% footer

%%%%%%%%%%%%%%%%%%%%%%%%%%%%%%%%%%%%%%%%%%%%%%%%%%%%%%
%%%%%%%%%%%%%%%%%% some definitions %%%%%%%%%%%%%%%%%%
%%%%%%%%%%%%%%%%%%%%%%%%%%%%%%%%%%%%%%%%%%%%%%%%%%%%%%

\def\and{\qquad\hbox{\small and}\qquad}

\def\hence{\qquad\hbox{\small hence}\qquad}
\def\with{\qquad\hbox{\small with}\qquad}
\def\kikezd{\parag\underbar}
\def\const{{\rm const}}

\def\bR{{\bf R}}
\def\bC{{\bf C}}

\def\cA{{\cal A}}
\def\cB{{\cal B}}
\def\cL{{\cal L}}
\def\cP{{\cal P}}
\def\cS{{\cal S}}
\def\cU{{\cal U}}

\def\smallcirc{{\raise 0.5pt \hbox{$\scriptstyle\circ$}}}
\def\smallover#1/#2{\hbox{$\textstyle{#1\over#2}$}}
\def\2{{\smallover 1/2}}
\def\4{{\smallover 1/4}}
\def\ccr{\cr\noalign{\medskip}}
\def\parag{\hfil\break}
\def\={\!=\!}
\def\D{
{D\mkern-2mu\llap{{\raise+0.5pt\hbox{\big/}}}\mkern+2mu}
}
\def\nablaslash{
{\nabla\mkern-2mu\llap{{\raise+0.5pt\hbox{\big/}}}\mkern+2mu}
}

\def\o{{\rm o}}
\def\O{{\rm O}}

\def\v0{\vec{0}}
\def\vb{\vec{b}}
\def\vc{\vec{c}}
\def\vJ{\vec{J}}
\def\vD{\vec{D}}
\def\vE{\vec{E}}
\def\vG{\vec{G}}
\def\vP{\vec{P}}
\def\vx{\vec{x}}
\def\vX{\vec{X}}
\def\vbeta{\vec\beta}

\def\vsigma{\vec\sigma}
\def\vnabla{\vec\nabla}

\def\barX{\overline{X}}

\def\semidirectproduct{
{\ooalign
{\hfil\raise.07ex\hbox{s}\hfil\crcr\mathhexbox20D}}
} %%%%% cf. TeX book p.356

%%%%%%%%%%%%%%%%%%%%%%%%%%%%%%%%%%%%%%%%%%%%%%%%%%%%%%
%%%%%%%%%%%%%%%%%%%%% numberings %%%%%%%%%%%%%%%%%%%%%
%%%%%%%%%%%%%%%%%%%%%%%%%%%%%%%%%%%%%%%%%%%%%%%%%%%%%%

\newcount\ch %%%%% ch(apters
\newcount\eq %%%%% eq(uations
\newcount\foo %%%%% foo(tnotes
\newcount\ref %%%%% ref(erences

\def\chapter#1{
\parag\eq = 1\advance\ch by 1{\bf\the\ch.\enskip#1}
}

\def\equation{
\eqno(\the\ch.\the\eq)\global\advance\eq by 1
}

\def\reference{
\parag\number\ref{.}\ \advance\ref by 1
}

\ch = 0 %%%%% global init ch(apter
%%%%% eq is set to 1 by \chapter
\foo = 1 %%%%% global init foo(tnote
\ref = 1 %%%%% global init ref(erence

%%%%%%%%%%%%%%%%%%%%%%%%%%%%%%%%%%%%%%%%%%%%%%%%%%%%%%
%%%%%%%%%%%%%%%%%%%%% the abstract %%%%%%%%%%%%%%%%%%%
%%%%%%%%%%%%%%%%%%%%%%%%%%%%%%%%%%%%%%%%%%%%%%%%%%%%%%

\title
\vskip 1.1cm
\authors
\vskip .23in

\parag
{\bf Abstract}

{\sl It is shown that the non-relativistic `Dirac' equation of
L\'evy-Leblond, we used recently to describe a spin $\2$ field
interacting  non-relativistically with a Chern-Simons gauge field,
can be obtained by lightlike reduction from $3+1$ dimensions.
This allows us to prove that the system is Schr\"odinger symmetric.
A spinor representation of the  Schr\"odinger group is presented.
Static, self-dual solutions,  describing spinor  vortices are given
and shown to be the non-relativistic limits of the fermionic vortices
found by  Cho et al. The construction is extended to external harmonic
and uniform magnetic fields.}

%%%%%%%%%%%%%%%%%%%%%%%%%%%%%%%%%%%%%%%%%%%%%%%%%%%%%%
%%%%%%%%%%%%%%%%%% PACS and all that %%%%%%%%%%%%%%%%%
%%%%%%%%%%%%%%%%%%%%%%%%%%%%%%%%%%%%%%%%%%%%%%%%%%%%%%


\noindent
43 pages

%\vskip13mm

%\noindent{dhp3.tex, Version 1.0 (\the\day/\the\month/\the\year)}

%\vskip13mm

\vfill\eject

%%%%%%%%%%%%%%%%%%%%%%%%%%%%%%%%%%%%%%%%%%%%%%%%%%%%%%
%%%%%%%%%%%%%%%%%%%%%%%%%%%%%%%%%%%%%%%%%%%%%%%%%%%%%%
\chapter{Introduction}
%%%%%%%%%%%%%%%%%%%%%%%%%%%%%%%%%%%%%%%%%%%%%%%%%%%%%%
%%%%%%%%%%%%%%%%%%%%%%%%%%%%%%%%%%%%%%%%%%%%%%%%%%%%%%

The `Chern-Simons' version of electro\-magnetism in $2+1$
dimensions, where the electromagnetic field, $(B,\vE)$,
and the current, $(\varrho,\vJ)$, satisfy the identities
$$
\kappa B\equiv\kappa\,\epsilon^{ij}\partial_iA^j=
-e\,\varrho,
\qquad
\qquad
\kappa E^i\equiv\kappa\left(-\partial_iA^0-\partial_tA^i\right)=
e\,\epsilon^{ij}J^j
\equation
$$
rather than the Maxwell equations, provides a promising framework for
understanding the quantized Hall effect~[1] as well as high $T_c$
superconductivity~[2].

The theory supports, just like the conventional Abelian Higgs model [3],
vortices [4]. Exact solutions were found, in particular, in the {\sl
non-relativistic} model of Jackiw and Pi~[5] who chose
$
\varrho=\tilde\Phi^*\tilde\Phi
$
and
$
\vJ
=
1/(2im)
\left[\tilde\Phi^*\vD\tilde\Phi-\tilde\Phi(\vD\tilde\Phi)^*\right]
$,
where $D_\alpha\equiv\partial_\alpha-ie\,A_\alpha$ (with
$\alpha=0,1,2$), the massive scalar field $\tilde\Phi$ satisfying the
planar, gauged, non-linear Schr\"o\-din\-ger (NLS) equation,
$$
iD_t\tilde\Phi=\left[
-{\vD^2\over 2m}
-\Lambda\,\tilde\Phi^*\tilde\Phi\right]\tilde\Phi.
\equation
$$
The model is non-relativistic: it admits the
Schr\"odinger group as symmetry [5], [6].
Let us stress the importance of such a development
for the physical applications:
condensed matter physics {\it is} non-relativistic.

For the special value
$\Lambda=e^2/(m\kappa)$ of the non-linearity,
the static second-order equation (1.2) can be reduced
to the first-order `self-duality' (SD) equations,
$$
\big(D_1\pm iD_2\big)\tilde\Phi=0.
\equation
$$
In a suitable gauge, this leads to Liouville's equation, and can
therefore be solved [5].

Motivated by the above mentioned potential applications,
recently [7] we generalized the Jackiw-Pi theory to
{\sl spinors}: we described a spin $\2$ non-relativistic
field by the $(2+1)$-dimensional version of the
non-relativistic `Dirac' equation proposed some
time ago by L\'evy-Leblond [8],
$$\left\{
\matrix{
(\vsigma\cdot\vD)\Phi
&+\hfill&2m\,\chi&=&0,
\ccr
D_t\,\Phi&+\hfill&i(\vsigma\cdot\vD)\chi\hfill&=&0,
\cr}\right.
\equation
$$
coupled to the Chern-Simons field (1.1) through
$$
\varrho=|\Phi|^2
\and
\vJ=i\big(\Phi^\dagger\vsigma\,\chi
-\chi^\dagger\vsigma\,\Phi\big).
\equation
$$
Here $\Phi$ and $\chi$ are two component \lq Pauli' spinors and
$\vsigma\cdot\vD\equiv\sum_{j=1}^2\sigma^jD_j$, with $\sigma^j$ denoting
the  Pauli matrices.
Then, Eq.~(1.4) is readily seen to split into two `chiral' components.
In a suitable
basis, the chiral components combine according to
$$
\Phi=\pmatrix{\Phi_+\cr\Phi_-},
\qquad
\qquad
\chi=\pmatrix{\chi_-\cr\chi_+},
\equation
$$
so that $\Phi$ and $\chi$ themselves are not chiral.
Eliminating the $\chi$-components  in Eq.~(1.4), we could finally
reduce our equations to
$$
iD_t\Phi_\pm=
\Big[-{\vD^2\over2m}
\pm\lambda\,\Phi_\pm^\dagger\Phi_\pm\Big]\Phi_\pm
\equation
$$
with
$$
\lambda\equiv{e^2\over2m\kappa},
\equation
$$
which is Eq.~(1.2) but with non-linearities {\sl half} of the
special value used by Jackiw and Pi in Ref.~[5].
Generalizing the SD equations (1.3) allowed us to construct
new, purely magnetic $(A_t=0$), spinorial vortex solutions [7]
of the coupled system (1.1,4,5).
The relation of the LHS of Eq.~(1.7) to the square of a first-order
planar system has already been noted [5].
Our result here provide a physical interpretation to this observation.

In another paper [6], we investigated the Jackiw-Pi model in
a `non-relativistic Kaluza-Klein' framework:
we started with a $(3+1)$-dimensional Lorentz manifold
$(M,g)$ endowed with a complete covariantly-constant null vector $\xi$,
called a `Bargmann space'~[9].
Non-relativistic spacetime, $Q$, is the quotient of
$M$ by the integral curves of $\xi$. On $M$, we posited
the relations
$$
\kappa f_{\mu\nu}=
e\sqrt{-g}\,\epsilon_{\mu\nu\rho\sigma}\,\xi^\rho j^\sigma
\qquad
\and
\qquad
\partial\mathstrut_{[\mu}f\mathstrut_{\nu\rho]}=0.
\equation
$$
The first of these equations --- we will call the FCI --- lifts the
field/current identity, Eq.~(1.1),  while the second guarantees
that $f_{\mu\nu}$ derives (locally) from a $4$-potential
$a_\mu$, i.e. $f_{\mu\nu}=\partial_\mu a_\nu-\partial_\nu a_\mu$.
The non-linear Schr\"odinger equation, Eq.~(1.2), is in turn obtained
from a non-linear wave equation for a massless scalar field, $\tilde\phi$,
on $M$, when $\tilde\phi$ is required to be  equivariant (with respect to
the group generated by $\xi$), viz
$
D_\xi\tilde\phi=im\tilde\phi.
$
As emphasized in Ref.~[6], this framework is particularly convenient
to describe
the {\sl symmetries} and the associated {\sl conserved quantities}
of the system.

In this paper, we extend our `Kaluza-Klein-type' framework to
{\sl spinors}.
In particular, we show that the L\'evy-Leblond (LL) equation (1.4) is
in fact the lightlike reduction of the {\sl massless Dirac equation}
on a simple example of a  Bargmann space~[10,11], namely flat Minkowski
space.
However, the lightlike dimensional reduction is more general and works for
{\sl any} Bargmann  space. In the general case, the conformal invariance
of the reduced system can be shown along the same lines as in [6]. We
demonstrate the usefulness of this `Kaluza-Klein' framework by
explicitly computing all the conserved quantities that belong to the
various residual conformal symmetries in the spinor Chern-Simons (CS) theory.
In addition to this, we also show how it enables one to  derive the spinor
representation of the Schr\"odinger group.

It is important to know some exact solutions of the theory. Here we
rederive the static, purely magnetic, self-dual spinor vortices, already
found in Ref.~[7] in a slightly novel way.

Finally, again motivated by the potential physical applications, we
extend our
theory to background harmonic oscillator and uniform magnetic fields.
We achieve
this by adapting the geometric  framework described in [12] to the
spinor theory and using it we present explicit exact solutions --- again
in the form of spinor vortices --- in these  background fields.

The simplest case of our theory (i.e. when the Bargmann space is just
flat Minkowski space)  is closely related to the model considered by
Cho et al. in Ref.~[13] where the CS gauge fields are coupled to
{\sl relativistic} spinor fields in $2+1$ dimensions. Here we
show that, within the Kaluza Klein framework, this model is obtained by
using a spacelike rather than a lightlike covariantly constant $\xi$
vector.
Furthermore, we show that our static vortices can be viewed in fact as
the {\sl non-relativistic limits} of the fermionic vortices of Cho et al.

Our paper is organized as follows: in Sec.~2 we present the coupled
gauged L\'evy-Leblond and Chern-Simons equations. Sec.~3 contains the
discussion of the conformal symmetry in the case of a general Bargmann
space.  In Sec.~4  we derive the spinor representation of the
Schr\"odinger group, while  in Sec.~5 we present the conserved quantities
belonging to the residual conformal  symmetries. In Sec.~6 the static
solutions are presented. Sec.~7 is devoted to the external field problem,
and in Sec.~8 we discuss the relation with the model considered
in Ref.~[13]. In Sec.~9 we briefly describe the  problem of the
L\'evy-Leblond equations in higher dimensions.  In Sec.~10 we collected
several  remarks and comments on the various possible extensions  of our
theory.  Finally, some technical computations needed to establish the
conformal invariance are collected in Appendix A.

%\vfill\eject

\goodbreak

%%%%%%%%%%%%%%%%%%%%%%%%%%%%%%%%%%%%%%%%%%%%%%%%%%%%%%
%%%%%%%%%%%%%%%%%%%%%%%%%%%%%%%%%%%%%%%%%%%%%%%%%%%%%%
\chapter{The coupled, gauged, L\'evy-Leblond
and Chern-Simons equations}
%%%%%%%%%%%%%%%%%%%%%%%%%%%%%%%%%%%%%%%%%%%%%%%%%%%%%%
%%%%%%%%%%%%%%%%%%%%%%%%%%%%%%%%%%%%%%%%%%%%%%%%%%%%%%

Let $(M,g,\xi)$ be a four-dimensional Bargmann manifold of signature
$(-,+,+,+)$, which is assumed to be endowed with a spin structure,
and choose Dirac matrices
$\gamma^\mu$ (with $\mu=0,1,2,3$) such that
$\gamma^\mu\gamma^\nu+\gamma^\nu\gamma^\mu=-2g^{\mu\nu}$.
Let $a_\mu$ be a ${\rm U}(1)$-gauge
potential on $M$ and denote by
$D_\mu\equiv\nabla_\mu-i e\,a_\mu$ the gauge and metric-covariant
derivative of a spinor fields.
The latter reads explicitly as
$$
\nabla_\mu\psi=\partial_\mu\psi-
\smallover1/8\left[
\gamma^\rho,\partial_\mu\gamma_\rho
-\Gamma^\sigma_{\mu\rho}\gamma_\sigma\right]\psi,
\equation
$$
with $\Gamma^\sigma_{\mu\rho}$ denoting the Christoffel symbols of the
metric.
Let us now consider
the {\sl gauged massless Dirac equation} on $M$,
$$
\D\psi\equiv\gamma^\mu D_\mu\psi=0.
\equation
$$
This equation is chiral-symmetric: the chirality operator, which is
defined using an ortho\-normal basis by
$$
\Gamma\equiv\gamma^5
=
-\smallover{{\sqrt{-g}}}/{4!}\,\epsilon_{\mu\nu\rho\sigma}
\gamma^\mu\gamma^\nu\gamma^\rho\gamma^\sigma,
\equation
$$
anticommutes with $\D$.
The $4$-component equation
(2.2) splits therefore into two un\-coupled $2$-component equations
for the two spinor fields $\psi_+$ and $\psi_-$, which are in fact
the chiral components, defined by
$$
\Gamma\psi_\pm=\mp i\psi_\pm.
\equation
$$
Our system of equations has to be supplemented by the chirality
constraint~(2.4).

The space of spinors carries a hermitian structure of signature
$(+,+,-,-)$, defined by
$
\overline{\psi}\equiv\psi^\dagger G
$,
where $G$ is determined by the requirements
$
\overline{\gamma}_\mu
=
G^{-1}\gamma_\mu^\dagger G=\gamma_\mu
$
and
$
G^\dagger=G
$.

For the metric-covariant derivative of the Dirac adjoint,
$\overline{\psi}$,the relation
$
\overline{\nabla_\mu\psi}=\nabla_\mu\overline{\psi}
$
holds. Therefore, $\D\overline{\psi}=0$ whenever $\psi$ solves
the Dirac equation (2.2).
It follows that the current
$$
j^\mu=\overline{\psi}\,\gamma^\mu\,\psi
\equation
$$
is conserved, $\nabla_\mu j^\mu=0$.
Then the coupled system (1.9) and (2.2,4,5) becomes self-consistent.

We now reduce the system from $4$ to $3$ dimensions.
We can choose on our four-dimensional Bargmann manifold $(M,g,\xi)$
a trivializing coordinate  system $(t,x,y,s)$ such that $\xi=\partial_s$.
Then the quotient, $Q$, of $M$ by the flow of $\xi$ is a Newton-Cartan
spacetime~[9] which is, hence, parametrized by non-relativistic time and
position, $t$ and $\vx=(x,y)$, respectively.

Owing to the FCI in (1.9),
the field strength $f_{\mu\nu}$ is the lift of a closed two-form
$F_{\alpha\beta}$ on $Q$.
The $4$-potential $a_\mu$ can be chosen therefore to be the
lift from $Q$ of a $3$-potential $A_\alpha$ (with $\alpha=0,1,2$)
such that $F_{\alpha\beta}=2\partial_{[\alpha}A_{\beta]}$.
As we demonstrated in Ref.~[6],
the lightlike reduction of the FCI in (1.9) yields precisely
the Jackiw-Pi field/current identity, Eq.~(1.1).
Similarly, the $4$-current $j^\mu$ projects onto the $3$-current
$J^\alpha=(\varrho,\vJ)$.

To reduce the spinor field we further require it to be
{\sl equivariant},
$$
D_\xi\psi=im\psi,
\equation
$$
for some constant $m$,
interpreted as the mass. Therefore, putting
$$
\psi=e^{ims}\pmatrix{\Phi\cr\chi},
\equation
$$
the two-component spinors $\Phi$ and $\chi$ do not depend on $s$ and
Eq.~(2.2) descends to one on $Q$.
Now it follows from the general theory of spinors that, in three
dimensions, `Dirac' spinors only have $2$ components.
This fact manifests itself in our theory in
that the reduction commutes with chirality.
Each of the two chiral components of Eq.~(2.2)
project therefore to two uncoupled systems in $2+1$ dimensions.
Identifying the chiral components with two-spinors,
we end up with two uncoupled, $2$-component equations.

In summary, the equations that describe the interacting
spinor and Chern-Simons gauge fields and the ones we intend to study
are the following:
$$
\D\psi=0,
\qquad
\qquad
\Gamma\psi=-i\varepsilon\psi,
\qquad
\qquad
D_\xi\psi=im\psi,
$$
$$
\kappa f_{\mu\nu}=
e\sqrt{-g}\,\epsilon_{\mu\nu\rho\sigma}\,\xi^\rho j^\sigma,
\qquad
\qquad
\partial_{[\mu}f_{\nu\rho]}=0,
\equation
$$
$$
j^\mu=\overline\psi\gamma^\mu\psi,
$$
with `helicity' $\varepsilon=\pm1$, mass $m$ and coupling
constants $\kappa$ and $e$.

\goodbreak

%%%%%%%%%%%%%%%%%%%%%%%%%%%%%%%%%%%%%%%%%%%%%%%%%%%%%%
\kikezd{Minkowski-space}
%%%%%%%%%%%%%%%%%%%%%%%%%%%%%%%%%%%%%%%%%%%%%%%%%%%%%%

Let us exemplify these general statements in the simplest case where
$M$ is Minkowski space, $\bR^{3,1}$, with its metric written in
light-cone coordinates as $d\vx^2+2dtds$.
The Dirac matrices
$$
\gamma^t=\pmatrix{0&0\cr1&0},
\qquad
\vec{\gamma}=\pmatrix{-i\vsigma &0\cr0&i\vsigma},
\qquad
\gamma^s=\pmatrix{0&-2\cr0&0},
\equation
$$
as well as
$$
\Gamma=
\pmatrix{-i\sigma_3&0\cr0&i\sigma_3},
\equation
$$
are hermitian,
$
\overline{\gamma^\mu}=\gamma^\mu,
$
with respect to the hermitian structure defined by
$
G=\pmatrix{0&1\cr1&0}.
$

The massless Dirac equation $\D\psi=0$, Eq.~(2.2), on $M$ reduces,
when equivariance, Eq.~(2.6), is imposed, to the first-order system
$$
\left\{
\matrix{
(\vsigma\cdot\vD)\Phi
&+\hfill&2m\chi&=&0,
\ccr
D_t\Phi&+\hfill&i(\vsigma\cdot\vD)\chi\hfill&=&0.
\cr}\right.
\equation
$$
which is {\sl formally} the same as the non-relativistic `Dirac'
equation of L\'evy-Leblond [8]
(albeit in $2+1$ rather than $3+1$ dimensions).
Note that
$\Phi=\pmatrix{\Phi_+\cr\Phi_-}$ and
$\chi=\pmatrix{\chi_-\cr\chi_+}$ are {\sl not}
the chiral components.
For $\Gamma$ as in (2.10), these latter read rather
$$
\Psi_+=\pmatrix{\Phi_+\cr0\hfill\cr0\hfill\cr\chi_+}
\and
\Psi_-=\pmatrix{0\hfill\cr\Phi_-\cr\chi_-\cr0\hfill},
\equation
$$
respectively.
Then, Eq.~(2.11) splits into the two uncoupled systems of prescribed
chirality
$$
\eqalign{
&\left\{\matrix{
&(D_1+iD_2)\Phi_+&+&2m\chi_+&=0,
\ccr
&D_t\Phi_+&+&i(D_1-iD_2)\chi_+&=0;\hfill
}\right.
\cr\cr
&\left\{\matrix{
&(D_1-iD_2)\Phi_-&+&2m\chi_-&=0,
\ccr
&D_t\Phi_-&+&i(D_1+iD_2)\chi_-&=0.\hfill
}\right.
}
\equation
$$

\goodbreak

As will be shown in Sec.~4, the $\psi_+$ and $\psi_-$ Bargmann spinor
fields
(see (2.7,12)) span the spin $\2$ and $-\2$ `spinor representations'
of the
Schr\"odinger group in $2+1$ dimensions, thus they are indeed `spinor'
fields.\foot{
Since Euclidian rotations form an Abelian subgroup of the Galilei
group in $2+1$ dimensions, spin is a subtle object. In fact, we
introduce spin by dimensional reduction so that the {\sl Poincar\'e
helicity} is  actually a Galilei invariant, whence the justification of
the notion of spin $\pm\2$.
}
Each of them separately would be sufficient to describe
(in general different) physical phenomena in $2+1$ dimensions.
Nevertheless, for the ease of presentation, we keep in the sequel
all four components of $\psi$, since in this way we can describe
the spin $\2$ and spin $-\2$ fields simultaneously.

It follows from the singular form of $\gamma^t$ in (2.9) that,
as found already by [8], the mass (or particle) density,
$\varrho\equiv J_\mu\xi^\mu$, viz
$\varrho=J_s= J^t$, only depends on the `upper' component,~$\Phi$:
$$
\varrho=|\Phi|^2=|\Phi_+|^2+|\Phi_-|^2
\equation
$$
and is positive definite allowing for a probabilistic interpretation.
For the spatial components of the current we find
$$
\vJ=i\big(\Phi^\dagger\vsigma\,\chi
-\chi^\dagger\vsigma\,\Phi\big).
\equation
$$
Combining Eqs~(2.14,15) with the FCI in (1.9) shows that the two
different
chirality spinors, if present simultaneously, couple to each other
only  through
the CS gauge fields.  Now $\chi$ can be eliminated using (2.11),
to yield  the
expression in terms of $\Phi$ only:
$$
\vJ={1\over2im}\Big(
\Phi^\dagger\vD\Phi-(\vD\Phi)^\dagger\Phi\Big)
+\vnabla\times\Big({1\over2m}\,\Phi^\dagger\sigma_3\Phi\Big).
\equation
$$
Note here the new term, due to the spin.

Using
$(\vsigma\cdot\vD)^2
=
\vD^2+eB\sigma_3
$,
we find that the component-spinors satisfy
$$
\left\{
\matrix{
&iD_t\Phi&=&-{1\over2m}\Big[\vD^2+eB\sigma_3\Big]\Phi,\hfill
\ccr
&iD_t\chi&=&-{1\over2m}\Big[\vD^2+eB\sigma_3\Big]\chi
-\smallover{e}/{2m}\,(\vsigma\cdot\vE)\,\Phi.
\cr}\right.
\equation
$$
Thus, $\Phi$ solves a `Pauli equation', while $\chi$ also couples
to the electric field through the Darwin term,
$\vsigma\cdot\vE$.

Expressing $\vE$ and $B$ through the FCI in (1.9) or (1.1)
and inserting them into our equations,
we get finally the remarkable system
$$
\left\{
\matrix{
&iD_t\Phi=
&\Big[-{1\over2m}\,\vD^2
+{e^2\over2m\kappa}\,|\Phi|^2\,\sigma_3
\Big]\Phi,\hfill
\ccr
&iD_t\chi=
&\Big[-{1\over2m}\,\vD^2
+\smallover{e^2}/{2m\kappa}\,|\Phi|^2\,\sigma_3
\Big]\chi
-\smallover{e^2}/{2m\kappa}\,\big(
\vsigma\times\vJ\big)\Phi.
\cr}\right.
\equation
$$
Note that the non-linearity comes from the Pauli term upon the
use of the FCI.

If we restrict the chirality of $\psi$ to $+1$ (respectively $-1$),
this system describes  non-relativistic spin $\2$ (spin $-\,\2$) fields
interacting  with a Chern-Simons gauge field.
Leaving the chirality of $\psi$ unspecified, the system  describes
{\sl two} spinor fields of spin $\pm\,\2$, interacting with
each other and the Chern-Simons gauge field.

Since the lower component of the spinor field is simply
$$
\chi=-{1\over2m}\,(\vsigma\cdot\vD)\Phi,
\equation
$$
it is enough to solve the
$\Phi$-equation.
In the general case this latter is still a coupled, $2$-component
equation,  because  $\varrho=|\Phi|^2$ involves both components of
$\Phi$.
For the $\pm$ chiral components in Eq.~(1.6), the `Pauli' equation for
$\Phi$ reduces to the equations (1.7) with the value (1.8) of the
non-linearity coupling constant.

\goodbreak

%%%%%%%%%%%%%%%%%%%%%%%%%%%%%%%%%%%%%%%%%%%%%%%%%%%%%%
%%%%%%%%%%%%%%%%%%%%%%%%%%%%%%%%%%%%%%%%%%%%%%%%%%%%%%
\chapter{Conformal symmetry}
%%%%%%%%%%%%%%%%%%%%%%%%%%%%%%%%%%%%%%%%%%%%%%%%%%%%%%
%%%%%%%%%%%%%%%%%%%%%%%%%%%%%%%%%%%%%%%%%%%%%%%%%%%%%%

In studying the symmetries of the NLS-CS equation, Ref.~[6],
we considered `non-relativistic conformal' transformation~[9,11],
i.e. those mappings $h:M\to{M}$ such that
$$
h^\star g_{\mu\nu}=\Omega^2g_{\mu\nu}
\and
h_\star\xi^\mu=\xi^\mu,
\equation
$$
for some smooth function $\Omega>0$.
Likewise, the non-relativistic conformal vector fields, $X$,
are such that, for some smooth function $k$, there holds
$$
L_Xg=k\,g
\qquad
\and
\qquad
L_X\xi=0.
\equation
$$

In this Section, we show that on any Bargmann space, the $\xi$-preserving
conformal transformations (3.1) still act as symmetries on  the coupled
spinor-CS system described by Eqs~(1.9) and (2.2,4,5,6)  --- or Eqs~(2.8).
The proof is given in  three steps: first we deal with  the conformal
invariance of the purely metric massless Dirac equation,
then  we prove the Bargmann conformal invariance of the ungauged
LL equations and, in the final step,  we consider the complete system
consisting of the gauged LL equations and the FCI.

\goodbreak

%%%%%%%%%%%%%%%%%%%%%%%%%%%%%%%%%%%%%%%%%%%%%%%%%%%%%%
\kikezd{Massless Dirac equation}
%%%%%%%%%%%%%%%%%%%%%%%%%%%%%%%%%%%%%%%%%%%%%%%%%%%%%%

Let us consider the purely metric Dirac operator,
$\nablaslash\equiv\gamma^\mu\nabla_\mu$, on a $(3+1)$-dimensional
Bargmann
manifold $(M,g,\xi)$ equipped with  Dirac matrices satisfying the
Clifford
relations $\gamma^\mu\gamma^\nu+\gamma^\mu\gamma^\mu=-2g^{\mu\nu}$,
where
$\nabla_\mu$ is the metric covariant derivative (2.1) of spinor fields.
To show the conformal invariance of $\nablaslash\psi=0$, we  recall that
the  Lie derivative of a spinor field with respect to a vector field $X$
is defined as~[14]
$$
L_X\psi
=
\nabla_X\psi
-\4
\gamma^\mu\gamma^\nu\partial_{[\mu}X_{\nu]}\psi.
\equation
$$
Let $X$ be an infinitesimal conformal transformation,
i.e. a vector field on $M$ such that
$$
L_X g_{\mu\nu}=2\nabla_{(\mu}X_{\nu)}=k\,g_{\mu\nu}
\equation
$$
for some function $k$ on $M$ (notice that $k=\2\nabla_\mu X^\mu$).

%\goodbreak

Evaluating the commutator $[L_X,\nablaslash]$ (see the details in
Appendix A) one finds
$$
[L_X,\nablaslash]\psi
=
-\smallover 1/2 k\,\nablaslash\psi+\smallover 3/4\gamma^\mu
\partial_\mu k\,\psi.
%\equation
$$
Therefore, defining the infinitesimal action of
conformal transformations on Dirac spinors to be rather
$$
\delta_X\psi\equiv L_X\psi+\smallover 3/8(\nabla_\mu X^\mu)\,\psi,
\equation
$$
one has
$
[\delta_X,\nablaslash]\psi
=
-\smallover 1/4(\nabla_\mu X^\mu)\,\nablaslash\psi
$.
Thus, whenever $\psi$ is a solution of the massless Dirac equation,
the same is true for
$$
\psi_\epsilon\equiv\psi+\epsilon\,\delta_X\psi+\cdots
\equation
$$
for any conformal-Killing vector field $X$ of $(M,g)$:
$$
\nablaslash\psi=0
\qquad
\Longrightarrow
\qquad
\nablaslash\delta_X\psi=0.
\equation
$$

\goodbreak

%%%%%%%%%%%%%%%%%%%%%%%%%%%%%%%%%%%%%%%%%%%%%%%%%%%%%%
\kikezd{The ungauged L\'evy-Leblond system}
%%%%%%%%%%%%%%%%%%%%%%%%%%%%%%%%%%%%%%%%%%%%%%%%%%%%%%

Next, we prove the invariance of the ungauged L\'evy-Leblond system
$$
\nablaslash\psi=0
\qquad
\and
\qquad
\nabla_\xi\psi=im\psi
\equation
$$
under the Lie algebra of the conformal-Killing vector fields
which also preserve the `vertical' vector field $\xi$, cf.~(3.2).

\goodbreak

Since the question of conformal invariance of the massless Dirac
equation has already been settled, one only has to prove that the
equivariance condition (or mass constraint) is preserved.
Clearly, $[L_X,L_\xi]=L_{[X,\xi]}=0$ entails
$
[\delta_X,L_\xi]\psi=0,
%\equation
$
as a result of the definitions and of
$
\xi^\mu\partial_\mu k=0.
$
Thus, if $\psi$ is a solution of the L\'evy-Leblond equation,
the same is true for $\psi_\epsilon$, Eq.~(3.6), for any
Bargmann-conformal vector field $X$, since
$$
L_\xi\psi=im\psi
\qquad
\Longrightarrow
\qquad
L_\xi\delta_X\psi=im\,\delta_X\psi.
\equation
$$

Using the expression of the Lie derivative of gamma matrices,
Eq.~(3.10) below, it is straightforward to prove that the chirality
condition (2.4) is also Bargmann-conformally invariant,
$$
[\delta_X,\Gamma]\psi=0.
$$

\goodbreak

%%%%%%%%%%%%%%%%%%%%%%%%%%%%%%%%%%%%%%%%%%%%%%%%%%%%%%
\kikezd{The gauged L\'evy-Leblond/Chern-Simons system}
%%%%%%%%%%%%%%%%%%%%%%%%%%%%%%%%%%%%%%%%%%%%%%%%%%%%%%

Let us now turn to our final goal, i.e., to prove the
Bargmann-conformal
invariance of the coupled, gauged, L\'evy-Leblond/Chern-Simons system,
expressed
on our extended `Bargmann' spacetime as Eqs~(2.8). It follows from the
definition (3.3) that
$$
L_X\gamma_\mu=\2\big(L_Xg_{\mu\nu}\big)\,\gamma^\nu.
\equation
$$
Hence, for any conformal-Killing vector field $X$,
$$
\delta_X\gamma_\mu
\equiv
L_X\gamma_\mu-\2 k\,\gamma_\mu=0.
\equation
$$
Putting
$\D\equiv\gamma^\mu D_\mu$ with
$D_\mu\equiv\nabla_\mu-ie\,a_\mu$, we find
$$
[\delta_X,\D]\psi
=
-ie\gamma^\mu(L_X a)_\mu\psi
-\smallover 1/2 k \D\psi,
\equation
$$
so that whenever $\psi$ and $a$ solve the gauged, massless, Dirac
equation (2.2), the same is true for $\psi_\epsilon$, Eq.~(3.6), and
$$
a_\epsilon\equiv a+\epsilon\,\delta_X a+\cdots
\equation
$$
i.e.
$$
\D\psi=0
\qquad
\Longrightarrow
\qquad
\D\delta_X\psi
-ie\gamma^\mu(\delta_X a)_\mu\psi
=0,
\equation
$$
{\sl provided} the CS-field  transforms as
$$
\delta_X a=L_X a,
\qquad
\hence
\qquad
\delta_X f\equiv L_X f.
\equation
$$

\goodbreak

The mass constraint, Eq.~(2.6), is also Bargmann-conformally
invariant: in fact,
$
[\delta_X,D_\xi]\psi
=
-ie(L_X a)(\xi)\psi,
%\equation
$
proving
$
L_\xi\delta_X\psi-ie(\delta_X a)(\xi)\psi=im\,\delta_X\psi,
%\equation
$
i.e. that, up to higher order terms,
$$
\big(\D_{\epsilon}\big)_\xi\psi_\epsilon=im\,\psi_\epsilon.
\equation
$$

The CS-field equations (1.9) are readily seen to be
Bargmann-conformally invariant since, under the transformations
(3.15), they change as
$$
\partial\mathstrut_{[\mu}\,\delta_X f\mathstrut_{\nu\rho]}=0
\and
\kappa\,\delta_X f_{\mu\nu}
=
e\sqrt{-g}\,
\epsilon_{\mu\nu\rho\sigma}\,\xi^\rho\delta_X j^\sigma,
\equation
$$
where $\delta_X j\equiv L_Xj+2k\,j$.
The proof is completed by establishing the consistency of these
equations,
using Eqs~(3.5) and (2.5) identifying the CS and Dirac currents:
$$
\delta_X\left(\overline\psi\gamma^\mu\psi\right)
=
\delta_X\overline\psi\gamma^\mu\psi
+
\overline\psi\gamma^\mu\delta_X\psi
=
L_X\left(\overline\psi\gamma^\mu\psi\right)
+ 2 k\,\overline\psi\gamma^\mu\psi.
$$

In conclusion, the Bargmann-conformal vector fields $X$ on $(M,g,\xi)$,
$$
\delta_X g\equiv L_X g-\2\left(\nabla_\mu X^\mu\right)g=0
\and
\delta_X\xi\equiv L_X\xi=0,
\equation
$$
provide symmetries of the coupled LL-CS Eqs, associated with the
representation (3.5) and (3.15) on the spinor and gauge fields,
respectively.
These vector fields form a finite dimensional Lie algebra that can be
integrated (see below) to a finite dimensional Lie group of
$\xi$-preserving conformal transformations.

\goodbreak

%%%%%%%%%%%%%%%%%%%%%%%%%%%%%%%%%%%%%%%%%%%%%%%%%%%%%%
%%%%%%%%%%%%%%%%%%%%%%%%%%%%%%%%%%%%%%%%%%%%%%%%%%%%%%
\chapter{
Spinor representation of the Schr\"odinger group}
%%%%%%%%%%%%%%%%%%%%%%%%%%%%%%%%%%%%%%%%%%%%%%%%%%%%%%
%%%%%%%%%%%%%%%%%%%%%%%%%%%%%%%%%%%%%%%%%%%%%%%%%%%%%%

In the case of Minkowski space $M=\bR^{3,1}$, those conformal vector
fields
which preserve $\xi=\partial_s$ (see (3.18)) form a $9$-parameter
Lie algebra,
called the (extended) `Schr\"odinger' algebra [15].
It  consists of the vector  fields
$
X
=
X^t{\partial_t}+
X^j{\partial_j}+
X^s{\partial_s}
$
(with $j=1,2$) of the form [9,11]
$$
(X^\mu)=\pmatrix{
\kappa t^2+2\chi\,t+\epsilon
\hfill
\ccr
\hat\omega\vx+t\vec{\beta}+\vec{\gamma}+(\chi+\kappa t)\vx
\hfill
\ccr
-\2\kappa\,r^2-\vec{\beta}\cdot\vx+\eta
\hfill
}
\equation
$$
with $r=||\vx||$ and
$\hat\omega_{ij}=\omega\epsilon_{ij}$ where $\omega\in\bR$;
$\vec{\beta},\vec{\gamma}\in\bR^2$;
$\epsilon,\kappa,\chi,\eta\in\bR$.

Integrating the infinitesimal action we get the (extended)
Schr\"odinger group [16,9] as a $9$-dimensional subgroup of
$\O(4,2)$ which acts on $M$ according to
$(t,\vx,s)\to(t^*,\vx{\,}^*,s^*)$:
$$
\left\{
\eqalign{
t^*&={\displaystyle{d t+e\over ft+g}}
\ccr
\vx{\,}^*&={\displaystyle{R\vx+\vb t+\vc\over ft+g}}
\ccr
s^*&={\displaystyle
s
+{f\over2}{\left(R\vx+\vb t+\vc\right)^2\over ft+g}
-\vb\cdot R\vx
-{t\over2}\vb{\,}^2+h
}
}
\right.
\equation
$$
while its induced action on spacetime, $Q=\bR\times\bR^2$,
is given by $(t,\vx)\to(t^*,\vx{\,}^*)$, i.e. by `forgetting'
about the $s$ coordinate in (4.2).
Here, $R\in{\rm SO}(2)$ is a rotation in the plane, and $\vb,\vc\in\bR^2$
are the boosts and the space translations, respectively;
at last $d,e,f,g\in\bR$ with $dg-ef=1$ parametrize the
${\rm SL}(2,\bR)$  subgroup formed by the time translations ($d=1,f=0$),
dilations ($e=f=0$), and expansions ($d=1,e=0$).
Denote by
$$
u=(a,\vb,\vc,d,e,f,g,h)
\equation
$$
a typical element of the so-called `Spin-Schr\"odinger' group where
$
a=\exp(\smallover i/2 \theta\sigma_3)
$
is such that
$
a(\vsigma\cdot\vx)a^{-1}=\vsigma\cdot(R\vx)
$.

We record, for further use, that {\sl dilations} and {\sl expansions}
form, indeed, a $2$-dimensional Lie subgroup consisting of
elements of the form
$(1,\vec{0},\vec{0},g^{-1}, 0,f,g,0)$ with $g>0$
and $f\in\bR$.
It will also be convenient to think of any element
(sitting in an open dense subset where $d=g^{-1}(1+ef)$)
of the Schr\"odinger group as the product of elements of
the $7$-dimensional Bargmann group (a central extension of the
Galilei group in $2+1$ dimensions) and of that $2$-dimensional group,
a Borel subgroup of ${\rm SL}(2,\bR)$, namely
$$
u
=
\left(
a,g\vb-f\vc,g^{-1}\vc,1,g^{-1}e,0,1,h+\2 fg^{-1}\vc\,{}^2
\right)
\cdot
\left(1,\vec{0},\vec{0},g^{-1},0,f,g,0\right).
\equation
$$

%\goodbreak

%%%%%%%%%%%%%%%%%%%%%%%%%%%%%%%%%%%%%%%%%%%%%%%%%%%%%%
\kikezd{The infinitesimal transformations}
%%%%%%%%%%%%%%%%%%%%%%%%%%%%%%%%%%%%%%%%%%%%%%%%%%%%%%

As previously shown, infinitesimal conformal transformations, $X$,
act on the space of solutions of the massless Dirac equation,
namely through the operators ${1\over{i}}\delta_X$ where
$$
\delta_X\psi
\equiv
\Big(X^\mu\nabla_\mu-\smallover1/8\big[\gamma^\mu,\gamma^\nu\big]
\nabla_\mu X_\nu
+\smallover3/8\nabla_\mu X^\mu\Big)\psi.
\equation
$$
In the Minkowski case, the infinitesimal conformal transformations form
the $\o(4,2)$ Lie algebra,
$$
\left\{
\matrix{
P_\mu&=&-i\partial_\mu\hfill
&\hbox{\small translations,}\hfill
\ccr
M_{\mu\nu}&=&-i\left(
x_\mu\partial_\nu-x_\nu\partial_\mu\right)
+\smallover{i}/4[\gamma_\mu,\gamma_\nu]\quad\hfill
&\hbox{\small Lorentz transformations},\hfill
\ccr
d&=&-ix^\mu\partial_\mu-\smallover3i/2\hfill
&\hbox{\small dilations},\hfill
\ccr
K_\mu&=&-i\left(
x_\nu x^\nu\partial_\mu
-x_\mu(3+2x^\nu\partial_\nu)\right)
-\smallover{i}/2[\gamma_\mu,\gamma_\nu]x^\nu\hfill
&\hbox{\small conformal transformations}.\hfill
\ccr}
\right.
\equation
$$
The Schr\"odinger algebra is precisely the subalgebra which leaves the
covariantly constant vector $\xi$ invariant.
The most convenient way of
finding it is to view $\xi$ as the generator of vertical translations,
$P_s=-i\partial_s$, and find those generators in $\o(4,2)$,
Eq.~(4.6),
which commute with it. These operators preserve also the equivariance
condition, Eq.~(2.6), as well as the chirality condition, Eq.~(2.4).
We find that Eq.~(4.5) reads in flat Minkowski spacetime
$$
\delta_X
\psi
=
X^\mu
\partial_\mu\psi
+
\pmatrix{
-i{\omega\over2}\sigma_3 -{1\over2}(\chi + \kappa t) \hfill & 0 \ccr
{i\over2}\vsigma\cdot(\vbeta+\kappa\vx) \hfill
&
\hfill -i{\omega\over2}\sigma_3 +{1\over2}(\chi + \kappa t)}
\psi
+
\smallover 3/2\,(\chi + \kappa t)
\psi
\equation
$$
where the Bargmann conformal vector fields $X$ are given by Eq.~(4.1).

\goodbreak

%%%%%%%%%%%%%%%%%%%%%%%%%%%%%%%%%%%%%%%%%%%%%%%%%%%%%%
\kikezd{The finite transformations}
%%%%%%%%%%%%%%%%%%%%%%%%%%%%%%%%%%%%%%%%%%%%%%%%%%%%%%

Now we derive the formul{\ae} for finite transformations
by integrating these expressions.
Firstly, we only consider dilations and expansions.

For expansions only, one gets from Eq.~(4.2)
$t^*=t/(ft+1)$, also $\vx{\,}^*=\vx/(ft+1)$
and $s^*=s+{1\over2}f\vx{\,}^2/(ft+1)$.
Some tedious calculation then
leads to the corresponding integration at the spinor level:
the general solution $f\to\psi_f$ of the differential equation
$d\psi/df=\delta_X\psi$
(with  all coefficients set zero in Eq.~(4.7)
with the exception of $\kappa=-1$)
reads
$$
\psi_f(t,\vx,s)
=
\pmatrix{
(ft+1)^{-1} \hfill & 0\ccr
{\displaystyle{f\over2i}\,(\vsigma\cdot\vx)(ft+1)^{-2}} \hfill
& (ft+1)^{-2}
}
\psi(t^*,\vx{\,}^*,s^*).
\equation
$$
Putting $\Psi(t,\vx)\equiv\psi(t,\vx,s)e^{-ims}$, one readily gets
$$
\Psi_{\!f}(t,\vx)
=
{1\over(ft+1)^2}
\pmatrix{
ft+1\hfill & 0\ccr
{\displaystyle{f\over2i}\,(\vsigma\cdot\vx)}\hfill & 1
}
\Psi\left({t\over ft+1},{\vx\over ft+1}\right)
\exp\left(-{im f\vx{\,}^2\over2(ft+1)}\right),
\equation
$$
and one verifies that we have, indeed, obtained a representation
$(\Psi_{\!f})_{f'}=\Psi_{\!f+f'}$ of the (additive) group of pure
expansions on the solutions of the LL equation.
As for pure dilations, a straightforward calculation yields
$$
\Psi_{\!d}(t,\vx)
=
\pmatrix{
d\hfill &  0\ccr
0\hfill &  d^2
}
\Psi(d^2t,d\,\vx),
\equation
$$
which, again, turns out to be a genuine representation.
Combining these two results by invoking the product of a dilation
and an
expansion $(d,0)\cdot(1,f)=(d,f/d)$), one then finds the following
(anti-)representation, $\pi$, of this $2$-dimensional Borel
subgroup
$$
\pi(u)\psi(t,\vx,s)
=
{1\over(ft+g)^2}
\pmatrix{
ft+g\hfill &  0\ccr
{\displaystyle{f\over2i}\,(\vsigma\cdot\vx)}\hfill & 1
}
\psi(t^*,\vx{\,}^*,s^*),
\equation
$$
where $(t^*,\vx{\,}^*,s^*)=u\cdot(t,\vx,s)$ is the image,
(4.2), by
$u=(1,\vec{0},\vec{0},g^{-1},0,f,g,0)$.
One directly checks that we have indeed an
anti-representation of the Borel subgroup, viz
$\pi(uv)=\pi(v)\pi(u)$.

On the other hand, a simple calculation gives the following
(anti-)representation, we still denote $\pi$, of the
{\sl Bargmann group}
(a mere group of isometries), originally due to L\'evy-Leblond~[8]
$$
\pi(u)\psi(t,\vx,s)
=
\pmatrix{
a^{-1}\hfill &  0\ccr
{\displaystyle{i\over2}\,a^{-1}\vsigma\cdot\vb}\hfill &a^{-1}
}
\psi(t^*,\vx{\,}^*,s^*),
\equation
$$
where $(t^*,\vx{\,}^*,s^*)=u\cdot(t,\vx,s)$ with
$u=(a,\vb,\vc,1,e,0,1,h)$ in the Spin-Bargmann group, see (4.3).

\goodbreak

We then use the previous factorization of the Schr\"odinger group to
define the sought representation, again called $\pi$, of the
Spin-Schr\"odinger
group  $\pi(u)=\pi(u'')\pi(u')$ associated with the decomposition
(4.4) of the
group element $u=u'u''$.
With the help of Eqs~(4.11,12), one ends up with
the following (anti-)representation of the full Spin-Schr\"odinger
group
$$
\pi(u)\psi(t,\vx,s)
=
{1\over(ft+g)^2}
\pmatrix{
a^{-1}(ft+g)\hfill &  0\ccr
{\displaystyle{1\over2i}\,
a^{-1}\vsigma\cdot(fR\vx-g\vb+f\vc)}\hfill &a^{-1}
}
\psi(t^*,\vx{\,}^*,s^*),
\equation
$$
where $(t^*,\vx{\,}^*,s^*)=u\cdot(t,\vx,s)$ with
$u=(a,\vb,\vc,d,e,f,g,h)$.

By equivariance, one finally obtains the `natural'
(anti-)representation of the full Spin-Schr\"odinger group on
the space of solutions  of the free LL equation, that is
$$
\eqalign{
\pi(u)\Psi&(t,\vx)
=
\ccr
&{1\over(ft+g)^2}
\pmatrix{
a^{-1}(ft+g)\hfill &0\ccr
{\displaystyle{1\over2i}\,
a^{-1}\vsigma\cdot(fR\vx-g\vb+f\vc)}\hfill &a^{-1}
}
\Psi\left({dt+e\over ft+g},{R\vx+\vb t+\vc\over ft+g}\right)
\ccr
&
\times
\exp\left(-im\left\{{f(R\vx+\vb t+\vc)^2\over2(ft+g)}-
\vb\cdot{R\vx}-{t\over2}\vb{\,}^2+h\right\}\right)
}
\equation
$$
where the group element $u$ is as in (4.3).

Observe that the action of the Schr\"odinger group commutes with that
of $\Gamma$, and descends therefore to the chiral components $\psi_\pm$.
This chiral representation is also of the `lower triangular' form:
though the $\Phi_\pm$ components are mapped into themselves,
the $\chi_+$ (resp.~$\chi_-$) components transform into combinations
of $\Phi_+$ and $\chi_+$  (resp.~$\Phi_-$ and $\chi_-$) under boosts
and expansion. It is, therefore, not possible to further reduce the
representation provided by the $\psi_\pm$.

\goodbreak

%%%%%%%%%%%%%%%%%%%%%%%%%%%%%%%%%%%%%%%%%%%%%%%%%%%%%%
%%%%%%%%%%%%%%%%%%%%%%%%%%%%%%%%%%%%%%%%%%%%%%%%%%%%%%
\chapter{Conserved quantities}
%%%%%%%%%%%%%%%%%%%%%%%%%%%%%%%%%%%%%%%%%%%%%%%%%%%%%%
%%%%%%%%%%%%%%%%%%%%%%%%%%%%%%%%%%%%%%%%%%%%%%%%%%%%%%

In Ref.~[6] we constructed, following Souriau~[17], a symmetric,
traceless, conserved energy-momentum tensor, $\vartheta_{\mu\nu}$,
for the NLS-CS system and used it to associate the conserved charge
$$
{\cal Q}_X=
\int\vartheta_{\mu\nu}X^\mu\xi^\nu\sqrt{\gamma}\,d^2\!\vx,
\equation
$$
where $\gamma=\det(g_{ij})$, to each $\xi$-preserving conformal
vector  field
$X$ of Bargmann space (here, the integral is taken over
`$2$-space at time $t$').
Now we extend and apply these results to spinors (see, e.g.~[18] and
references therein). Remember that the massless Dirac equation can be
obtained by varying the {\sl matter action}
$$
\cS
=
\int_M\Im\big\{\overline{\psi}\,\D\psi\big\}\sqrt{-g}\,d^4\!x.
\equation
$$
We note that the variational derivative with respect to the vector
potential, $\delta\cS/\delta a_\mu$, yields the current $j^\mu$ in
Eq.~(2.5), whose conservation, $\nabla_\mu j^\mu=0$, follows simply
from the invariance of $\cS$ with respect to gauge transformations.

Denote $\cL$ the {\sl Lagrangian density} in definition (5.2),
namely $\cS=\int_M\cL$. Then a short calculation resorting to the action
of a conformal vector field, $X$, on Dirac spinors, Eq.~(3.5), and on the
CS field, Eq.~(3.15), together with the technical results (3.11,12),
shows that
$$
\delta_X\cL
=
-\2 k\,\cL + L_X\cL
\equation
$$
where $k=\2\nabla_\mu X^\mu$ (see (3.18)).
Now $\cL=0$ in view of the Euler-Lagrange equations (2.2),
it thus follows, since $\cL$ is represented by a closed $4$-form,
that
$
\delta_X\cL=0
$
modulo a surface term that stems from the Lie derivative in the RHS of
Eq.~(5.3). Thus, for any conformal vector field $X$, we have
$$
\delta_X\cS=0
\equation
$$
at the critical points of the matter action, $\cS$, which again proves
the
conformal symmetry of our system.
The energy-momentum tensor
$\vartheta_{\mu\nu}=-2\,\delta\cS/\delta g^{\mu\nu}$, viz
$$
\vartheta_{\mu\nu}
=-2\,{\delta\cS\;\;\over\delta g^{\mu\nu}}
=\2\,\Im\Big\{
\overline{\psi}\big(\gamma_\mu D_\nu+\gamma_\nu D_\mu\big)\psi\Big\},
\equation
$$
is therefore traceless as well as conserved and automatically symmetric.
Hence, a conserved quantity, Eq.~(5.1), is associated
to each $\xi$-preserving conformal vector field.
Upon using Eqs~(5.1,5) and (4.1) as well as
the reduced Dirac equation, Eq.~(2.11), and
assuming all surface terms vanish,
the associated conserved quantities read
$$
\left\{
\matrix{
\vP=&\displaystyle\int
\underbrace{\left\{
{1\over2i}
\left(\Phi^\dagger\vD\Phi-(\vD\Phi)^\dagger\Phi\right)
\right\}}_{\vec\cP}\,d^2\!\vx\hfill
&\hbox{\small linear momentum,}\hfill
\ccr
M
=&
m\displaystyle\int|\Phi|^2\,d^2\!\vx
\hfill&\hbox{\small mass $\times$ particle number,}\hfill
\ccr
J=&\underbrace{\displaystyle\int
\vec{r}\times\vec\cP\,d^2\!\vx
}_{{\small orbital}}
\;+\;
\underbrace{
\2\displaystyle\int\Phi^\dagger\sigma_3\Phi\,d^2\!\vx
}_{{\small spin}}
\hfill
&\hbox{\small angular momentum,}\hfill
\ccr
\vG=&t\vP
-m\displaystyle\int|\Phi|^2\,\vx\,d^2\!\vx
\hfill
&\hbox{\small boost,
}\hfill
\ccr
H
=&
\displaystyle\int\left\{
{1\over2m}|\vD\Phi|^2
+\lambda
|\Phi|^2\Phi^\dagger\sigma_3\Phi
\right\}\,d^2\!\vx\;
\hfill
\hfill&\hbox{\small energy,}\hfill\ccr
D=&2tH-\displaystyle\int\vec\cP\cdot\vx\,d^2\!\vx
\hfill
&\hbox{\small dilation,
}\hfill
\ccr
K=&-t^2H + tD + {m\over2}\displaystyle\int
|\Phi|^2\,\vx{\,}^2\,d^2\!\vx
\hfill
&\hbox{\small expansion.
}\hfill
\cr
}
\right.
\equation
$$
where the constant $\lambda=e^2/(2m\kappa)$ is as in (1.8).

\goodbreak

%%%%%%%%%%%%%%%%%%%%%%%%%%%%%%%%%%%%%%%%%%%%%%%%%%%%%%
\kikezd{A $(2+1)$-dimensional formulation}
%%%%%%%%%%%%%%%%%%%%%%%%%%%%%%%%%%%%%%%%%%%%%%%%%%%%%%

As pointed out in Sec.~2, when the massless Dirac equation is reduced
to $2+1$ dimensions, the resulting equations split, consistently with
chiral symmetry, into two independent sets of equations.
This can be seen also by looking at the action.
Identifying (with some abuse of notation) $\Psi_\pm$ with
two-component spinors,
$$
\Psi_+=\pmatrix{\Phi_+\cr\chi_+}
\and
\Psi_-=\pmatrix{\Phi_-\cr\chi_-},
\equation
$$
and introducing the two sets of $2\times2$ matrices
$$
\Sigma^t_\pm=\2(1+\sigma_3),
\qquad
\Sigma^1_\pm=-\sigma_2,
\qquad
\Sigma^2_\pm=\pm\sigma_1,
\qquad
\Sigma^s_\pm=\2(1-\sigma_3)
\equation
$$
taking into account the mass constraint, the Lagrangian density
in Eq.~(5.2) is seen in fact to split into two parts,
$$
\Im\left\{
\psi_+^\dagger
\big(\Sigma^t_+D_t+\Sigma^i_+D_i-2im\Sigma^s_+\big)\psi_+
\right\}
+
\Im\left\{
\psi_-^\dagger
\big(\Sigma^t_-D_t+\Sigma^i_-D_i-2im\Sigma^s_-\big)\psi_-
\right\},
\equation
$$
whose variational equations are
$$
\Big[
\Sigma^t_\pm D_t+\vec\Sigma_\pm\cdot\vD-2im\Sigma^s_\pm
\Big]\Psi_\pm=0,
\equation
$$
i.e. the L\'evy-Leblond equations (2.13), in another form.
Augmenting (5.9) by the Lagrangian density of the CS gauge fields,
$$
\cL_{CS}
=
{\kappa\over4}\epsilon^{\alpha\beta\gamma}A_\alpha F_{\beta\gamma},
\equation
$$
we get a {\sl local $(2+1)$-dimensional} Lagrangian for the complete
spinor-CS system.
Thus, just like in the scalar case, while there is no action for
the coupled system in $3+1$ dimensions, in $2+1$ dimensions there
is one.

The existence of the `reduced' action is important for at least
two reasons:
on the one hand, it makes it possible to  apply the standard canonical
or Hamiltonian description to the spinor-CS system, and on the other,
it opens up an alternative way to discuss the symmetries and the
conserved quantities.
These investigations go beyond the scope of the present paper and
here we merely give the  canonical Hamiltonian density:
$$
{\cal H}
=
A_t\left(e\sum\limits_{i=\pm}\psi_i^\dagger\Sigma^t_i\psi_i
 +\kappa F_{12}\right)
+
2m\sum\limits_{i=\pm}\psi_i^\dagger\Sigma^s_i\psi_i-
\Im\left\{\sum\limits_{i=\pm}\psi_i^\dagger
\vec{\Sigma_i}\cdot(\vD\psi_i)\right\}.
\equation
$$
The static equations of motion can also be obtained by variation of
the Hamiltonian. It is easy to see --- using the equations of motions ---
that on {\sl solutions} $\int{\!{\cal H}\,d^2\!\vx}$ takes
(up to an irrelevant sign)
the same value as $H$, Eq.~(5.6).

\goodbreak

%\vfill\eject

%%%%%%%%%%%%%%%%%%%%%%%%%%%%%%%%%%%%%%%%%%%%%%%%%%%%%%
%%%%%%%%%%%%%%%%%%%%%%%%%%%%%%%%%%%%%%%%%%%%%%%%%%%%%%
\chapter{Static, self-dual solutions}
%%%%%%%%%%%%%%%%%%%%%%%%%%%%%%%%%%%%%%%%%%%%%%%%%%%%%%
%%%%%%%%%%%%%%%%%%%%%%%%%%%%%%%%%%%%%%%%%%%%%%%%%%%%%%

%%%%%%%%%%%%%%%%%%%%%%%%%%%%%%%%%%%%%%%%%%%%%%%%%%%%%%
\kikezd{Reduction to the Liouville equation}
%%%%%%%%%%%%%%%%%%%%%%%%%%%%%%%%%%%%%%%%%%%%%%%%%%%%%%

In Ref.~[7], we constructed static, purely magnetic vortices of
definite chirality by solving the static versions of Eqs~(2.17)
and (1.1) using the  first-order `self-dual' Ansatz
$$
(D_1\pm iD_2)\Phi=0.
\equation
$$
Here we arrive at this same equation in a more direct  way.
Let us consider in
fact the static version of the first-order equations (2.13),
written in chiral components:
$$
\eqalign{
&\left\{\matrix{
&(D_1+iD_2)\Phi_+&+&2m\chi_+&=0,
\ccr
&-e\,A_t\Phi_+&+&(D_1-iD_2)\chi_+&=0;
\cr}\right.
\ccr
&\left\{\matrix{
&(D_1-iD_2)\Phi_-&+&2m\chi_-&=0,
\ccr
&-e\,A_t\Phi_-&+&(D_1+iD_2)\chi_-&=0.
\cr}\right.
\cr}
\equation
$$
Then, requiring the solution to be purely magnetic, $A_t=0$,
and setting $\chi_+=0$ for the positive and $\chi_-=0$ for the
negative chirality,
these equations reduce directly to the self-duality conditions (6.1).
In this sense, the first-order LL equations (2.13) already
contain self-duality.  It is readily seen that the current,
Eq.~(2.15),
vanishes identically for positive chirality spinors with $\chi_+=0$
(negative chirality ones with $\chi_-=0$), thus with $A_t\equiv 0$ the
static  version of the Chern-Simons equations,
$\kappa E^i=e\epsilon^{ij}J^j$, are also satisfied.

The SD conditions (6.1), as well as the remaining Chern-Simons
equation, $B=-e\varrho/\kappa$, where $\varrho=|\Phi_+|^2$
(or $|\Phi_-|^2$) were solved in [7] along the lines of Ref.~[5]:
in the gauge $\Phi_\pm=\varrho^{1/2}$,\foot{
The gauge transformation needed to bring $\Phi_\pm$ to this form
may be singular at the zeros of $\varrho$.
}
Eq.~(6.1) leads to
$$
\vec{A}
=
\pm{1\over2e}\vnabla\times\ln\varrho.
\equation
$$
Thus $B=-e\varrho/\kappa$ reduces, in both cases, to the
Liouville equation
$$
\bigtriangleup\ln\varrho=\pm{2e^2\over\kappa}\varrho.
\equation
$$
A normalizable solution is obtained for
$\Phi_+$ when $\kappa<0$, and for $\Phi_-$ when $\kappa>0$.
These correspond precisely to having an attractive non-linearity
in Eq.~(2.19).

\goodbreak

%%%%%%%%%%%%%%%%%%%%%%%%%%%%%%%%%%%%%%%%%%%%%%%%%%%%%%
\kikezd{The properties of the solutions}
%%%%%%%%%%%%%%%%%%%%%%%%%%%%%%%%%%%%%%%%%%%%%%%%%%%%%%

As these static solutions have only the $\Phi_+$ or $\Phi_-$
components of
$\Phi$  not identically zero, they involve only
{\sl one} of the spinor fields $\Psi_\pm$ in $2+1$ dimensions (see
(2.12)),
depending  upon the sign of
$\kappa$.

\goodbreak

Furthermore, as the total magnetic flux of the solutions,
$$
\int{\!B\,d^2\!\vx}
=
-{e\over\kappa}\int{\!\varrho\,d^2\!\vx}
\equiv
-{e\over\kappa}N,
\equation
$$
is nonvanishing if $N\neq0$,
we call them spinor vortices. Also, as $A_t\equiv 0$
(and hence $\vE=0$), they are {\sl purely magnetic} ones.

We can now confirm our results on self-duality.
The same  argument as in the scalar case [5] shows that
static solutions must have vanishing energy:
it is the only possibility to cancel the apparent $t$-dependence in
the conserved quantities $D$ (dilation) and $K$ (expansion).
Using the algebraic identity:
$
|\vD\Phi|^2=|(D_1\pm iD_2)\Phi|^2 \pm\varrho
$
(up to total derivatives integrating to surface terms), which is
established along the same lines as in the scalar case [5], the energy,
Eq.~(5.7), is re-written as
$$
H={1\over2m}\int\big|(D_1\pm iD_2)\Phi\big|^2\,d^2\vx
\mp 2\lambda\int\vert\Phi_\mp\vert^2\,\Phi^\dagger\Phi\,d^2\vx,
\equation
$$
where $\Phi=\pmatrix{\Phi_+\cr\Phi_-}$.
Now it is easy to see that this expression indeed {\sl vanishes}
for the (anti) self dual configurations described above:
taking the upper (resp. lower) sign for $\kappa<0$
(resp. $\kappa>0$)  we get a positive semidefinite expression, and
$\Phi_-\equiv 0$  (resp. $\Phi_+\equiv 0$) together with the first
(resp. second)  equation in
(6.1) guarantee that it is indeed zero.

It is straightforward to verify that the {\sl static}
field equations (6.1) can also be obtained by variation of
$\int{\!{\cal H}\,d^2\!\vx}$.
However, as,
in contrast to the scalar case [5],
${\cal H}$ is {\sl not} a (semi) definite expression, we can
not rule out the existence of non self-dual static solutions.

The general solution of the Liouville equation is
$$
\varrho=\mp{4\kappa\over e^2}{|f'(z)|^2\over(1+|f(z)|^2)^{2}}
\equation
$$
with $z=x^1+ix^2$ and $f(z)$ complex analytic, but for our purposes
$f(z)$
must be chosen such that $\varrho$ is well behaved.
To describe explicit self-dual solitons we use the radially
symmetric solution of Eq.~(6.7), given by [5]:
$$
\varrho_n(r)=\mp{4n^2\kappa\over e^2r^2}\left[
\left({r_0\over r}\right)^n
+
\left({r\over r_0}\right)^n
\right]^{-2}
\equation
$$
where $r=|z|$, with $r_0$ and $n$ two free parameters.
However, as explained in [5],
the aforementioned (singular) gauge transformation, which not
only removes the phases of $\Phi_\pm$, but also cancels the
singularities in $B$, must be single valued.
This condition requires $n$
to be an integer, which, without loss of generality, can be chosen
to be
positive. Integrating $\varrho_n$ over $2$-space yields
$$
N={\vert\kappa\vert\over e^2}4\pi n.
\equation
$$
This quantity then determines the actual values of all
the conserved charges for our solitons: both the magnetic flux,
$-eN/\kappa=-({\rm sign}\,\kappa/e)4\pi n$,
and the mass, $M=mN$,
are the same as for the scalar soliton of [5], while the total
angular momentum, $J=\mp N/2$ is merely {\sl half} of the
corresponding value taken  by the scalar soliton
(the conserved quantities $\vP,\vG$ and $D$ vanish for our spinor
solitons much in the same way as they do for the scalar one).

\goodbreak

%\vfill\eject

%%%%%%%%%%%%%%%%%%%%%%%%%%%%%%%%%%%%%%%%%%%%%%%%%%%%%%
%%%%%%%%%%%%%%%%%%%%%%%%%%%%%%%%%%%%%%%%%%%%%%%%%%%%%%
\chapter{Spinor vortices in external fields}
%%%%%%%%%%%%%%%%%%%%%%%%%%%%%%%%%%%%%%%%%%%%%%%%%%%%%%
%%%%%%%%%%%%%%%%%%%%%%%%%%%%%%%%%%%%%%%%%%%%%%%%%%%%%%

Physical applications as the Fractional Quantum Hall Effect
[1] would require to extend the spinor theory to background fields.
In the scalar case, the `empty' space solution has been
`exported' to provide solutions in a uniform magnetic or harmonic
force field [19].
In Ref.~[12] we gave a geometric framework to this procedure
and showed that the only cases where it works correspond
to the time-dependent
uniform magnetic and electric or harmonic forces.
Now we generalize
this approach for the case of spinors.

To describe the external fields one can use the most general
`Bargmann' metric found long ago by Brinkmann [20]:
$$
g_{ij}(t,\vx) dx^idx^j+2dt\left[ds+{\cal A}_i(t,\vx)dx^i\right]
-2\cU(t,\vx)dt^2,
\equation
$$
where the `transverse' metric $g_{ij}$ (with $i,j=1,2$) as well
as the `vector potential' $\vec{\cal A}$ and the `scalar potential'
$\cU$ are functions of $t$ and $\vx$ only.
Clearly, $\xi=\partial_s$ is a covariantly constant null vector.
The null geodesics of this metric describe particle motion
in curved transverse space in external electromagnetic fields
$\vec{\cal E}\sim-\partial_t\vec{\cal A}-\vnabla\cU$
and $\cB\sim\vnabla\times\vec{\cal A}$ [9].
In the sequel, we make the simplifying assumption that the transverse
$2$-space is {\sl flat}, $g_{ij}=\delta_{ij}$.

Consider now the Dirac-Chern-Simons system, Eqs~(2.8), in the
background (7.1).
Using that the non-vanishing components of the inverse metric are
$g^{ij}=\delta^{ij}$,
$g^{is}=-\cA^i$,
$g^{ss}=2\cU+\cA_i\cA^i$,
$g^{ts}=g^{st}=1$,
we find that an appropriate set of `curved space' Dirac matrices can
be written as
$$
\gamma^t=\gamma^T,
\qquad
\gamma^i=\gamma^I,
\qquad
\gamma^s
=-\cA_i\gamma^I+\cU\gamma^T+\gamma^S,
\equation
$$
where  $\gamma^T$, $\gamma^I$ and $\gamma^S$ denote the constant, flat
space Dirac matrices (2.9):
$$
\gamma^T=\pmatrix{0&0\cr1&0},
\qquad
\gamma^I=\pmatrix{-i\sigma^I&0\cr0&i\sigma^I},
\qquad
\gamma^S=\pmatrix{0&-2\cr0 &0}.
\equation
$$
Since the determinant of the metric, hence $\sqrt{-g}$ as well as
$\gamma^i$ and $\gamma^t$ coincide with their flat Minkowski space
counterparts, both the $(3+1)$-dimensional and the  reduced
versions of the FCI, Eq.~(1.9) and Eq.~(1.1) respectively, retain their
previous, flat space form.
Nevertheless the external potentials, $\cA_i$ and $\cU$, do make their
presence felt through the Dirac equation.
The {\sl first} term in the massless Dirac equation (2.2):
$$
\gamma^\mu{\cal D}_\mu\psi
\equiv
\gamma^\mu D_\mu\psi
-\smallover1/8\left[
\gamma^\rho,\partial_\mu\gamma_\rho
-\Gamma^\sigma_{\mu\rho}\gamma_\sigma\right]\psi
=
0,
\equation
$$
(where $D_\mu\equiv\partial_\mu-ie\,a_\mu$) reads, if we put
$\psi\equiv e^{ims}\Psi$ and use the Dirac matrices (7.2),
$$
\Big(
\gamma^T(D_t+im\cU)+\gamma^J(D_j-im\cA_j)+im\gamma^S
\Big)\Psi.
%\equation
$$
Thus, the effect of these terms is simply to modify the
flat space CS-covariant
derivatives $D_\mu$ into `external field' ones, ${\cal D}_\mu$, viz
$$
{\cal D}_t=D_t-im\cA_t,
\qquad
{\cal D}_j=D_j-im\cA_j,
%\equation
$$
where $\cA_t=-\cU$.
Using the Christoffel symbols of the metric (7.1),
$$
\eqalign{
\Gamma^i_{tt}&=\partial_t\cA_i+\partial_i\cU,\ccr
\Gamma^i_{jt}&=\2\left(\partial_j\cA_i-\partial_i\cA_j\right),\ccr
\Gamma^s_{ij}&=\2\left(\partial_i\cA_j+\partial_j\cA_i\right),\ccr
\Gamma^s_{it}&=
-\partial_i\left(\cU+\4\vec\cA{\,\strut}^2\right)
+\2\cA^j\partial_j\cA_i,\ccr
\Gamma^s_{tt}&=
-\cA^i\partial_i\cU
-\partial_t\left(\cU+\2\vec\cA{\,\strut}^2\right),
}
%\equation
$$
one finds that the net contribution of the
{\sl second} term to the massless Dirac equation (7.4) has the form:
$$
-{1\over16}
\left(\partial_i\cA_j-\partial_j\cA_i\right)
[\gamma^I,\gamma^J]\gamma^T\Psi.
%\equation
$$
Collecting these terms, we get  (without imposing the chirality
constraint) that the reduced massless Dirac equation takes the form:
$$
\left\{
\matrix{
(\vsigma\cdot\vec{\cal D})\Phi&&+&&2m\chi&=&0,\ccr
{\cal D}_t\Phi&+&{i\over4}\cB\sigma_3\Phi
&+&i(\vsigma\cdot\vec{\cal D})\chi\hfill&=&0,
\cr}\right.
\equation
$$
where $\cB=\epsilon^{ij}\partial_i\cA_j$ denotes
the background `magnetic' field.

\goodbreak

Eliminating the lower component we see that $\Phi$ satisfies the
`Pauli equation' with anomalous magnetic moment (compare (2.18):
$$
i{\cal D}_t\Phi=
-{1\over2m}\left[\vec{\cal D}{\,}^2+eB\sigma_3\right]\Phi
-{\cB\over4}\sigma_3\Phi.
\equation
$$

In order to identify the `magnetic' component of our metric
($\cA_j$) with the potential of a  genuine magnetic field,
one has to make the replacement $\cA_j\to e\,\cA_j/m$.
Then the $\cB$-dependent term in Eqs~(7.5,6),
$e\cB/(4m)\sigma_3\Phi$, shows that our reduced spinor fields
--- since they have a non vanishing magnetic moment
--- interact also directly
with the background magnetic field, in addition to the minimal
coupling appearing in ${\cal D}_\mu$. For chiral spinors
the sign of the magnetic moment depends
on the sign of the $4$-dimensional chirality.

%%%%%%%%%%%%%%%%%%%%%%%%%%%%%%%%%%%%%%%%%%%%%%%%%%%%%%
\kikezd{The harmonic oscillator field}
%%%%%%%%%%%%%%%%%%%%%%%%%%%%%%%%%%%%%%%%%%%%%%%%%%%%%%

Physically, the most interesting external fields describe either a
uniform
magnetic field or a harmonic oscillator force field. The latter case is
obtained by choosing
$
\cA_i=0
$
and
$
\cU=\2\omega^2r^2
$
in (7.1) [12], so
that, in the oscillator background, the Dirac matrices are:
$$
\gamma^t=\gamma^T,
\qquad
\gamma^i=\gamma^I,
\qquad
\gamma^s=\2\omega^2r^2\,\gamma^T+\gamma^S.
\equation
$$
The clue that makes possible to obtain the solutions of the oscillator
problem from that of the Minkowski space, is that the
mapping $(t,\vx,s)\to(T,\vX,S)\equiv(\tilde x^\mu)$:
$$
T={\tan\omega t\over\omega},\qquad X^i={x^i\over\cos\omega t},\qquad
S=s-{\omega r^2\over2}\tan\omega t,
\equation
$$
carries the oscillator metric
$$
d\vx{\,}^2+2dtds-\omega^2r^2dt^2,
\equation
$$
conformally into the flat metric $d\vX{\,}^2+2dTdS$ [12].
The sought connection between the two sets of solutions
can be determined by implementing this conformal mapping
on the spinors.
This can be done in three steps:
invoking the equivariance condition, one writes the solution of the
oscillator
problem  as
$$
\psi=\exp\left(im\left(s-\2\omega r^2\tan\omega t\right)\right)
\tilde\Psi(T,\vX)
$$
and substitutes it into Eq.~(7.4) with  Dirac
matrices as in Eq.~(7.7). In this way, one gets
$$
\exp\left(im\left(s-\2\omega r^2\tan\omega t\right)\right)
\left[\tilde\gamma^t{\cal D}_T+\tilde\gamma^j{\cal D}_J
+im\tilde\gamma^s\right]
\tilde\Psi(T,\vX)=0,
\equation
$$
where
$$
\eqalign{
\tilde\gamma^t
&=
\gamma^T{1\over\cos^2\omega t},
\ccr
\tilde\gamma^i
&=
\gamma^T\,{x^i\omega\sin\omega t\over\cos^2\omega t} +
\gamma^I\,{1\over\cos\omega t},
\ccr
\tilde\gamma^s
&=
\gamma^S-\gamma^Ix_i\omega\tan\omega t-\gamma^T\,
{\omega^2r^2\over2}\tan^2\omega t.
}
\equation
$$
It is straightforward to verify that $\tilde\gamma^\mu$ satisfy the
anticommutation relations
$$
\tilde\gamma^\mu\tilde\gamma^\nu + \tilde\gamma^\nu\tilde\gamma^\mu
=
-2\tilde{g}^{\mu\nu}=
-{2\over\cos^2\omega t}\,\eta^{\mu\nu}
\equation
$$
where $\eta^{\mu\nu}$ denotes the flat Minkowski metric. This equation
implies that $\tilde{g}_{\mu\nu}=\eta_{\mu\nu}\cos^2\omega t$, therefore
as a second step we make a conformal rescaling by
$\Omega=\cos\omega t$. Now using that
$\tilde{\cal D}\tilde\Psi=\Omega^{-5/2}\hat {\cal D}\hat\Psi$ if
$\tilde\Psi=\Omega^{-3/2}\hat\Psi$  and
adding to this the observation that
$\hat\gamma^\mu=\tilde\gamma^\mu\cos\omega t$ satisfy the anticommutation
rules of flat Minkowski space,  one can rewrite Eq.~(7.10) as
$$
\exp\left(im\left(s-\2\omega r^2\tan\omega t\right)\right)
{(\cos\omega t)^{-5/2}\over\cos\omega t}
\left[\hat\gamma^t{\cal D}_T+\hat\gamma^j{\cal D}_J+
im\hat\gamma^s\right]
\hat\Psi(T,\vX)
=
0.
\equation
$$
The next step is to find operators $U$ and $U^{-1}$ that `rotate'
$\hat\gamma^\mu$ into the flat space Dirac matrices:
$$
\hat\gamma^t=U\gamma^TU^{-1},
\qquad
\hat\gamma^i=U\gamma^IU^{-1},
\qquad
\hat\gamma^s=U\gamma^SU^{-1}.
\equation
$$
Making a suitable lower triangular Ansatz for $U$ and exploiting the
explicit form of the flat space Dirac matrices, one readily finds that
$$
U^{-1}=f\pmatrix{
1&0\ccr
-i{\omega\over2}(\vsigma\cdot\vx)\sin\omega t&\cos\omega t
},
\qquad
U=f^{-1}\pmatrix{
1&0\ccr
i{\omega\over2}(\vsigma\cdot\vx)\tan\omega t&(\cos\omega t)^{-1}
}
$$
do this for {\sl any} $f$. The most natural choice is to require
$\det{U}=1$, this yields then $f=(\cos\omega t)^{-1/2}$.
Since ${\cal D}_\mu$ is a `covariant derivative' we can write
$$
\hat\gamma^\mu{\cal D}_\mu\hat\Psi
=
U\gamma^\mu U^{-1}{\cal D}_\mu\hat\Psi
=
U\gamma^\mu{\cal D}_\mu(U^{-1}\hat\Psi)
$$
in Eq.~(7.13). Therefore, denoting the solution of the original
L\'evy-Leblond equation by $\Psi^0(T,\vX)$, the solution of the
oscillator problem can be written as
$$
\Psi(t,\vx)
=
\exp\left(-im{\omega r^2\over2}\tan\omega t\right)
(\cos\omega t)^{-3/2}
U\Psi^0(T,\vX),
$$
hence
$$
\eqalign{
\Psi(t,&\vx)=\hfill\cr
&\exp\left(-im{\omega r^2\over2}\tan\omega t\right)
(\cos\omega t)^{-1}
 \pmatrix{
1&0\ccr
i{\omega\over2}(\vsigma\cdot\vx)\tan\omega t
&(\cos\omega t)^{-1}\cr}\Psi^0(T,\vX).
}
\equation
$$
Note the two interesting properties of this final expression:

(a) The various powers of $\cos\omega t$ combine nicely to yield
a pure $-1$ for
the upper component, $\Phi$, of $\Psi^0$; i.e. $\Phi$ transforms
with the same
`time-dependent dilation' factor as a scalar field. In the same
time the `lower'
component, $\chi$, has a conformal factor $(\cos\omega t)^{-2}$
as well as an
inhomogeneous part, which is linear in $\Phi$.

(b)  Applying the transformation (7.15) to the static spinor
vortices in
Sec.~6 yields in particular time-dependent background-field solutions
with
non-vanishing `lower' component as well as a (Chern-Simons)
electric field.

%%%%%%%%%%%%%%%%%%%%%%%%%%%%%%%%%%%%%%%%%%%%%%%%%%%%%%
\kikezd{Constant magnetic field}
%%%%%%%%%%%%%%%%%%%%%%%%%%%%%%%%%%%%%%%%%%%%%%%%%%%%%%

The constant magnetic field background can be described by setting
$$
\cA_i=-
{e\over2m}\epsilon_{ij}\cB\,x^j,
\qquad
\cU= 0
\equation
$$
in (7.1); here we
already implemented the $\cA_j\to e\,\cA_j/m$ scaling. The key point
is to realize that the mapping (which amounts to switching
to a rotating frame with angular velocity $\omega=e\cB/(2m)$):
$(t,\vx,s)\to(t_{\rm osc},\vx_{\rm osc},s_{\rm osc})$
given by
$$
t_{\rm osc}=t,
\qquad
\vx_{\rm osc}=R(-\omega t)\vx,
\qquad
s_{\rm osc}=s,
\equation
$$
takes the `constant $\cB$-metric' (7.16) into the oscillator metric
(7.9) [12]. Just as in Sec.~4. the matrix $R(\theta)$ is a rotation by
angle $\theta$ in the plane and
$\vsigma\cdot(R\vx)=a(\vsigma\cdot\vx)a^{-1}$
with
$a(\theta)\equiv\exp(i\theta\sigma_3/2)$.
Repeating the same procedure as in the case of the oscillator
problem one finds that the transition  between the uniform $\cB$
field and the oscillator can be implemented by
$$
\Psi^\cB(t,\vx)=
\pmatrix{
e^{i\omega t\sigma_3/2}\hfill
&0
\ccr
-i{\omega\over2}
(\vsigma\times\vx)e^{i\omega t\sigma_3/2}
\quad\hfill
&e^{i\omega t\sigma_3/2}\hfill
}
\Psi^{\rm osc}(t,\vx_{\rm osc}).
\equation
$$
Combining the two conformal mappings in Eqs~(7.8) and (7.17)
gives finally a map from the uniform $\cB$ metric to the flat one [7]:
$$
T={\tan\omega t\over\omega},
\qquad
\vX={1\over\cos\omega t}\,R^{-1}(\omega t)\,\vx,
\equation
$$
and using the explicit expressions to implement these
transformations on the spinors, Eqs~(7.15) and (7.18),
finally we get~[7]:
$$
\eqalign{
\Psi^\cB&(t,\vx)=\hfill\ccr
&
{e^{-(im\omega r^2\tan\omega t)/2}\over\cos\omega t}\,
\pmatrix{
e^{i\omega t\sigma_3/2}\hfill
&0
\ccr
i{\omega\over2}\big[\tan\omega t
(\vsigma\cdot\vx)
-(\vsigma\times\vx)\big]e^{i\omega t\sigma_3/2}
\quad\hfill
&{e^{i\omega t\sigma_3/2}\over\cos\omega t}\hfill
}
\Psi^0(T,\vX),
\cr\cr
&A^\cB_\alpha=\partial_\alpha X^\beta A^0_\beta.
\qquad
\cr}
\equation
$$

%%%%%%%%%%%%%%%%%%%%%%%%%%%%%%%%%%%%%%%%%%%%%%%%%%%%%%
%%%%%%%%%%%%%%%%%%%%%%%%%%%%%%%%%%%%%%%%%%%%%%%%%%%%%%
\chapter{
Relativistic fermions and the non-relativistic limit
}
%%%%%%%%%%%%%%%%%%%%%%%%%%%%%%%%%%%%%%%%%%%%%%%%%%%%%%
%%%%%%%%%%%%%%%%%%%%%%%%%%%%%%%%%%%%%%%%%%%%%%%%%%%%%%

In Ref.~[13], Cho et al. consider the {\sl massive} gauged Dirac
equation
$
\D\psi=iM\psi
$
on four-dimensional Minkowski space with metric given by
$$
d\ell^2
=
-c^2d{t}^2+d{x}^2+d{y}^2+d{z}^2
\equation
$$
and require that the chiral components, denoted by
$\psi_\pm$, satisfy
$$
\psi_\pm({t},{x},{y},{z})
=
e^{ip_\pm{z}}\Psi_\pm({t},{x},{y}).
\equation
$$
Then their massive Dirac equation becomes, with $c=1$,
$$
\big(i\gamma^\alpha_\pm D_\alpha-p_\pm\big)\Psi_\pm
=
\pm M\sigma_3\Psi_\mp,
\equation
$$
where $\alpha=0,1,2$ and
$
(\gamma^\alpha_\pm)=(\pm\sigma^3, i\sigma^2,-i\sigma^1).
%\equation
$
Requiring the gauge field dynamics to be governed by the
Chern-Simons rule,
$$
\2\kappa\,\epsilon^{\alpha\beta\gamma}F_{\beta\gamma}
=
e\big(\overline{\Psi}_+\gamma^\alpha_+\Psi_+
+ \overline{\Psi}_-\gamma^\alpha_-\Psi_-\big),
\equation
$$
allows them to construct purely magnetic solutions.
They note also that for
$
p_+=p_-
$
and for chiral spinors with one vanishing component,
their equations reduce formally to the same system as the one
studied by Jackiw and Pi [5].
This is the point we will be discussing in this Section,
together with the relation between our work and that of Ref.~[13].

Setting  $M=0$ in Eq.~(8.3) results in two {\sl relativistic}  Dirac
equations on $\bR^{2,1}$ if one interprets $p_\pm$ as the masses of the
chiral components $\Psi_\pm$. Apart from the `mass' term $M$ (that
destroys conformal invariance), the main difference of Eq.~(8.3) with
our approach is that the reduction here is {\sl spacelike}, while we
used so far a lightlike reduction.
To proceed, we thus outline the relativistic version of our theory,
namely the coupled Dirac-Chern-Simons system (2.8) where, this time,
$$
g_{\mu\nu}\,\xi^\mu\xi^\nu=c^{-2}.
\equation
$$

%%%%%%%%%%%%%%%%%%%%%%%%%%%%%%%%%%%%%%%%%%%%%%%%%%%%%%
\kikezd{Relativistic spinors in Kaluza-Klein framework}
%%%%%%%%%%%%%%%%%%%%%%%%%%%%%%%%%%%%%%%%%%%%%%%%%%%%%%

Let us start with $\bR^{3,1}$, endowed with the metric [10]
$$
d\ell^2\equiv
g_{\mu\nu}dx^\mu dx^\nu
=
2dtds+dx^2+dy^2+c^{-2}ds^2
\equation
$$
where $c={\rm const}<+\infty$.
The (covariantly) constant vector field
$$
\xi={\partial\over\partial s}
\equation
$$
is chosen as the generator of vertical translations.
Note that $\xi$ is now spacelike, Eq.~(8.5).

\goodbreak

The contravariant metric associated with (8.6) descends as a
contravariant metric on the quotient
$\bR^{2,1}=\bR^{3,1}\!/\bR\xi$ which is plainly
Minkowski spacetime, with the metric $-c^2dt^2+dx^2+dy^2$.

Consider, as before but with $c<+\infty$,
the gauged, massless, Dirac equation on~$\bR^{3,1}$
$$
\gamma^\mu D_\mu\psi=0,
\equation
$$
with Dirac matrices
$$
\gamma^t=\pmatrix{0&c^{-2}\cr1&0},
\qquad
\vec{\gamma}=\pmatrix{-i\vec{\sigma}&0\cr0&i\vec{\sigma}},
\qquad
\gamma^s=\pmatrix{0&-2\cr0 &0},
\equation
$$
which indeed satisfy
$\gamma^\mu\gamma^\nu+\gamma^\nu\gamma^\mu=-2g^{\mu\nu}$; these are
again hermitian with respect to the metric $G=\pmatrix{0&1\cr1&0}$ on the
spinor space, $\bC^{2,2}$, as is the chirality operator
$$
\Gamma=\pmatrix{-i\sigma_3&0\cr0&i\sigma_3}
\equation
$$
which retains the same form as in (2.10).
The latter still anticommutes with the Dirac operator,
and also commutes with reduction with respect to the mass constraint
$$
\xi^\mu D_\mu\psi=im\psi.
\equation
$$

The massless Dirac equation (8.8), as well as its reduction,
splits  therefore
into two sets of prescribed chirality,
$$
\Gamma\psi_\pm=\mp i\psi_\pm.
\equation
$$
With the same assumption as in Sec.~2 and with the notation (2.7), the
projection of our system onto spacetime,
$\bR^{2,1}$, reads now
$$
\left\{\eqalign{
&c^{-2}\,i D_t\chi
+
\big(\vsigma\cdot\vD\big)\Phi+2m\chi=0,
\ccr
&D_t\Phi+i\big(\vsigma\cdot\vD\big)\chi=0,
\cr}\right.
\equation
$$
while the chiral components are still given by (2.12).
Note here the new term with $c^{-2}$.

The gauge field dynamics is again governed
by the CS field equations (1.9) which project as~[6]
$$
\2\kappa\,\epsilon^{\alpha\beta\gamma}F_{\beta\gamma}
=
e J^\alpha
\qquad
\and
\qquad
\partial_{[\alpha}F_{\beta\gamma]}=0
\equation
$$
(with $\alpha,\beta,\gamma=0,1,2$). The current in Eq.~(2.5),
descends as
$$
J^\alpha=
e\overline{\Psi}\gamma^\alpha\Psi
\equation
$$
so that $\varrho\equiv{J^t}=|\Phi|^2+c^{-2}|\chi|^2$ is still positive
definite and $\vJ$ is as in (2.15).

\goodbreak

%%%%%%%%%%%%%%%%%%%%%%%%%%%%%%%%%%%%%%%%%%%%%%%%%%%%%%
\kikezd{Symmetry}
%%%%%%%%%%%%%%%%%%%%%%%%%%%%%%%%%%%%%%%%%%%%%%%%%%%%%%

Equations (8.13--15) stem from the relativistic system (2.8)
in $3+1$ dimensions, whose invariance with respect to the
$\xi$-preserving conformal transformations had been established in
Sec.~3.
Now we show that these transformations span merely the
{\it trivial} extension
$$
e(2,1)\times\bR
\equation
$$
of the Poincar\'e Lie algebra $e(2,1)$ by the vertical translations
generated by $\xi$.

To prove this, remember that the algebra $\o(4,2)$
is spanned by those vector fields
$$
Z^\mu
=
\Lambda^\mu_{\ \nu} x^\nu
+\Gamma^\mu
+K^\mu x_\nu x^\nu-2 x^\mu K_\nu x^\nu+\alpha x^\mu
\equation
$$
where $\Lambda\in\o(3,1)$ (Lorentz transformation),
$\Gamma, K\in\bR^{3,1}$
(translation, special conformal transformation) and
$\alpha\in\bR$ (dilation).
Demanding $L_Z\xi=0$ results, since $\xi$ is (covariantly)
constant, in the extra condition
$
\nabla_\xi Z=0.
$
Since $Z^\mu$ is a quadratic expression of the coordinates, this
condition yields, for any $x^\mu$,
$$
K^\mu (x_\nu\xi^\nu)-x^\mu (K_\nu\xi ^\nu)
-(K_\nu x^\nu)\xi^\mu=0
\and
\Lambda^\mu_{\ \nu}\xi^\nu+\alpha\xi^\mu=0.
$$
It follows from the first relation that $K_\nu\xi^\nu=0$ and
$K^\mu=a\,\xi^\mu$ for some $a\in\bR$.  Using that $\xi$ is
non-null, we have $a=0$, hence $K=0$: the special conformal
transformations are  broken by the reduction.
Similarly, the second relation gives
$
\Lambda^\mu_{\ \nu}\,\xi_\mu\xi^\nu
=
\alpha\,\xi_\mu\xi^\mu=0
$
since $\Lambda_{\mu\nu}$ is skew-symmetric,
therefore $\alpha=0$: the dilations are broken and
the Lorentz transformations preserve the `vertical' vector $\xi$.
Thus, since $\xi$ is spacelike, $\Lambda$ sits in
$\o(2,1)$.
In conclusion, the residual symmetry is spanned by
$$
Z^\mu=\Lambda^\mu_{\ \nu} x^\nu+\Gamma^\mu
\qquad
\with
\qquad
\Lambda^\mu_{\ \nu}\xi^\nu=0.
\equation
$$
These vector fields form the $7$-dimensional Lie algebra
$
\o(2,1)\semidirectproduct\bR^{3,1}
$
which is $e(2,1)\times\bR$.

We thus contend that the system (8.13--15) admits the
relativistic symmetry~(8.16).

%%%%%%%%%%%%%%%%%%%%%%%%%%%%%%%%%%%%%%%%%%%%%%%%%%%%%%
\kikezd{Relation to the work of Cho et al.}
%%%%%%%%%%%%%%%%%%%%%%%%%%%%%%%%%%%%%%%%%%%%%%%%%%%%%%

In order to exhibit the relation of our relativistic system to that of
Ref.~[13], introduce the new coordinate system
$$
\tilde{t}=t,
\qquad
\tilde{x}=y,
\qquad
\tilde{y}=-x,
\qquad
\tilde{z}=s/c+ct
\equation
$$
in terms of which the metric (8.6) and the vertical vector
$\xi$, Eq.~(8.7), respectively read
$$
d\ell^2
=
-c^2 d\tilde{t}^2+d\tilde{x}^2+d\tilde{y}^2+d\tilde{z}^2
\and
\xi={1\over c}{\partial\over\partial\tilde{z}},
\equation
$$
and correspond precisely to those used in Ref.~[13]
(see Eqs~(8.1,2)).
The associated metric on $\bR^{2,1}=\bR^{3,1}\!/\bR\xi$,
which is again $(2+1)$-dimensional Minkowski spacetime,
is $-c^2 d\tilde{t}^2+d\tilde{x}^2+d\tilde{y}^2$.

The Dirac matrices in this new coordinate system are given by
$$
\tilde\gamma^t=\pmatrix{0&c^{-2}\cr1&0},
\;
\tilde\gamma^x=\pmatrix{-i\sigma_2&0\cr0&i\sigma_2},
\;
\tilde\gamma^y=\pmatrix{i\sigma_1&0\cr0&-i\sigma_1},
\;
\tilde\gamma^z=\pmatrix{0&-c^{-1}\cr c &0},
\equation
$$
and the chirality operator remains unchanged (compare Eq.~(8.10)),
$$
\tilde\Gamma
=
\pmatrix{-i\sigma_3&0\cr0&i\sigma_3}.
\equation
$$

\goodbreak

Now, the key point is that our Dirac-CS system (2.8) --- with
(8.5) --- is {\sl intrinsic}; as such, it can be projected upon
$(2+1)$-dimensional spacetime using the equivariance relation (8.11)
and the fact that the CS field strength is basic, independently of
any coordinate system. The former reads
$
\tilde\xi^\mu\tilde{D}_\mu\tilde\psi=im\tilde\psi
$
(where
$
\tilde{D}_\mu
\equiv
{\partial/\partial\tilde{x}^\mu}-ie\,\tilde{a}_\mu
$),
whence
$$
\tilde\psi(\tilde{t},\tilde{x},\tilde{y},\tilde{z})
=
e^{imc\tilde{z}}\tilde{\Psi}(\tilde{t},\tilde{x},\tilde{y})
\equation
$$
in a gauge where
$
\tilde{a}_\mu d\tilde{x}^\mu
$ is basic and, in view of (8.12,22),
$$
\tilde\Psi
=
\pmatrix{
\tilde\Phi_+\cr
\tilde\Phi_-\cr
\tilde\chi_-\cr
\tilde\chi_+}.
\equation
$$
The gauged, massless, Dirac equation
$
\tilde\gamma^\mu\tilde{D}_\mu\tilde\psi=0
$
then yields the couple of equations
$$
\pmatrix{
\tilde{D}_t + im c & \tilde{D}_1-i\tilde{D}_2\ccr
-(\tilde{D}_1+i\tilde{D}_2) & -c^{-2}\tilde{D}_t + imc^{-1}
}
\pmatrix{
\tilde\Phi_+\ccr
\tilde\chi_+
}
=
0
\equation
$$
and
$$
\pmatrix{
-c^{-2}\tilde{D}_t + imc^{-1} & \tilde{D}_1-i\tilde{D}_2\ccr
-(\tilde{D}_1+i\tilde{D}_2) &\tilde{D}_t + im c
}
\pmatrix{
\tilde\chi_-\ccr
\tilde\Phi_-
}
=
0.
\equation
$$
Upon putting $c=1$ and defining
$$
\tilde\Psi_+
=
\pmatrix{
\tilde\Phi_+\ccr
\tilde\chi_+
}
\qquad
\and
\qquad
\tilde\Psi_-
=
\pmatrix{
\tilde\chi_-\ccr
\tilde\Phi_-
},
\equation
$$
this system, Eqs~(8.25,26), is interestingly rewritten as
$$
\big(i\tilde\gamma^\alpha_\pm\tilde{D}_\alpha -m\big)\tilde\Psi_\pm=0
\equation
$$
where $\alpha=0,1,2$ and $(\tilde\gamma^\alpha_\pm)=(\pm\sigma^3,
i\sigma^2,-i\sigma^1)$, i.e. precisely the equations of Cho et al. with
the same chiral masses,
$p_\pm=mc$ (no parity violation) and with $M=0$ (see (8.3)).

\goodbreak

At last, tedious but straightforward calculations confirm that our
projected current
$$
\tilde{J}^\alpha
=
\overline{\tilde\Psi}\tilde\gamma^\alpha\tilde\Psi
\equation
$$
(see (8.15)) actually reproduces the current which appears in~(8.4),
$$
\tilde{J}^\alpha
=
\overline{\tilde\Psi}_+\tilde\gamma^\alpha_+\tilde\Psi_+
+ \overline{\tilde\Psi}_-\tilde\gamma^\alpha_-\tilde\Psi_-,
\equation
$$
where is is understood that the hermitian structures on the chiral
spinors (8.27) in this last expression are respectively
$G_+=\pmatrix{1&0\cr0&-1}$ and
$G_-=\pmatrix{-1&0\cr0&1}$.

%%%%%%%%%%%%%%%%%%%%%%%%%%%%%%%%%%%%%%%%%%%%%%%%%%%%%%
\kikezd{The non-relativistic limit}
%%%%%%%%%%%%%%%%%%%%%%%%%%%%%%%%%%%%%%%%%%%%%%%%%%%%%%

The coordinate system (8.6) is particularly suited for studying the
non-relativistic limit $c\to+\infty$ of the previous system. In these
coordinates, the constant vector field $\xi$ remains unchanged when
$c\to+\infty$, but tends to a null vector as the metric tends to the
canonical flat Bargmann metric $2dtds+dx^2+dy^2$. The Dirac matrices
also tend smoothly to their lightlike form (2.9) used in the
non-relativistic theory. Thus, both the massless Dirac equation and the
equivariance constraint reduce to the ones we used in Sec.~2.

Our system is hence the non-relativistic limit of that of Cho et al.
To explain this in slightly different way, observe that, in terms of the
projected quantities, our transformation formula (8.23) relates Cho et
al.'s spinor fields to ours, via the familiar high frequency oscillating
factor
$$
\tilde\Psi(\tilde{t},\tilde{x},\tilde{y})
=
e^{-imc^2t}\,\Psi(t,x,y).
\equation
$$
This results in transforming Eqs (8.25,26) into ours, Eq. (8.13).
Then taking the non-relativistic limit amounts simply to dropping
the term $c^{-2}D_t\chi$, which leaves us with the L\'evy-Leblond
equation (1.4).

Also, the $7$-dimensional relativistic symmetry algebra
$e(2,1)\times\bR$ (see (8.16)) gets enlarged into the
$9$-dimensional `Schr\"odinger' algebra (4.1).
In fact, for $\xi$ lightlike, one special conformal transformation
remains unbroken: $K^\mu=a\xi^\mu$ yields precisely the expansion
$a\in\bR$. Similarly, in the relation
$\Lambda^\mu_{\ \nu}\xi^\nu+\alpha\xi^\mu=0$
the dilation $\alpha\in\bR$ remains arbitrary. The relativistic dilation
combines, hence, with a Lorentz transformation to yield the
non-relativistic dilation in Eq.~(4.1).

Let us stress that, in the limit $c\to+\infty$,
all geometric structures and equations remain well
behaved on $\bR^{3,1}$
--- unlike on the $(2+1)$-dimensional quotient where singularities
occur.

\goodbreak

%%%%%%%%%%%%%%%%%%%%%%%%%%%%%%%%%%%%%%%%%%%%%%%%%%%%%%
\kikezd{Self-duality}
%%%%%%%%%%%%%%%%%%%%%%%%%%%%%%%%%%%%%%%%%%%%%%%%%%%%%%

Relativistic, purely-magnetic fermionic vortex solutions were constructed
by Cho et al. in Ref.~[13].  Remarkably, their solutions are again
associated with the Liouville equation, just as in (6.5). Let us
finish this Section by explaining how this comes about.
Observe that, for static and purely magnetic fields
$$
D_t\Psi=\big(\partial_t-ieA_t\big)\Psi=0.
\equation
$$

\goodbreak

Then the gauged, massless, Dirac equation (8.13) and the associated
SD equations retain the same form in either the relativistic
or non-relativistic regime, because the term involving $c^{-2}$ vanishes
along with $D_t\chi$.

\goodbreak

%%%%%%%%%%%%%%%%%%%%%%%%%%%%%%%%%%%%%%%%%%%%%%%%%%%%%%
%%%%%%%%%%%%%%%%%%%%%%%%%%%%%%%%%%%%%%%%%%%%%%%%%%%%%%
\chapter{
The L\'evy-Leblond equation in higher dimensions}
%%%%%%%%%%%%%%%%%%%%%%%%%%%%%%%%%%%%%%%%%%%%%%%%%%%%%%
%%%%%%%%%%%%%%%%%%%%%%%%%%%%%%%%%%%%%%%%%%%%%%%%%%%%%%

Our non relativistic Kaluza-Klein framework can also be applied
to  derive the
field equations of the non  relativistic spinor fields in $3+1$
dimensions. One
starts with a  {\sl $5$-dimensional} Bargmann space, $M$.
In $5$ dimensions  the
spinor fields, $\psi$, have $2^{[5/2]}=4$ components,
and in the absence of a
non trivial analogue of the chirality operator, $\Gamma$, they form
--- even in the massless case ---
a single irreducible representation of $\o(4,1)$.\foot{
This $\o(4,1)$ is the Lie algebra of the symmetry group of
the tangent space of $M$. It is a subalgebra of the $\o(5,2)$ Lie
algebra formed by the infinitesimal conformal transformations of $M$.
}
Since the FCI, Eq.~(1.9), have no obvious analogue in $5$ dimensions,
$\psi$ can only couple  to the `standard' non-relativistic
Maxwell electrodynamics, described, e.g., in [9].
If $M$ is flat Minkowski space with metric
$\sum_{j=1}^3(dx^j)^2+2dtds$,then  we can choose the Dirac matrices
$$
\gamma^t=\pmatrix{0&0\cr1&0},
\qquad
\gamma^j=\pmatrix{-i\sigma^j&0\cr0&i\sigma^j},
\qquad
\gamma^s=\pmatrix{0&-2\cr0 &0}
\equation
$$
with $j=1,\ldots,3$.
Using these $\gamma^\mu$ in the massless Dirac equation
$\D\psi=0$, together with the mass constraint, Eq.~(2.6),
leads to a system having the same form as
Eq.~(2.11),  the only difference being is that now
$\vsigma\cdot\vD=\sum_{j=1}^3\sigma^jD_j$.
This system is identical to that proposed by L\'evy-Leblond [8].

The absence of the analogue of $\gamma^5$ implies that this reduced
system does
not now  split into two independent spinor equations,
although the $\Phi$ part
can again be solved separately.

The fact that $M$ has now $5$ dimensions changes slightly the
discussion of conformal invariance and the description of the
spinor representation of the Schr\"odinger group.
In particular
$$
\widehat{\D}\widehat\psi=\Omega^{-3}\,{\D}\psi
\with
\widehat{\psi}=
\Omega^{-2}\psi,
\equation
$$
so that the current scales as
$j^\mu\to\widehat{\jmath}{\,}^\mu=\Omega^{-5}j^\mu$.
Also, the coefficient of the last term in Eq.~(4.5) changes
from $3/8$ to $7/16$.
As a consequence, some numerical factors in the expressions of
$d$ and $K_\mu$ in (4.6) change  (e.g. the $3/2$ in $d$ changes
to $35/16$), but basically this equation
(with $\mu$ running over $t,s$ and $1,2,3$) determines the action
of the generators of $\o(5,2)$ on the solutions of $\D\psi=0$.
Naively, this seems to be a contradiction, since $\o(5,2)$
has no four-dimensional representation. Indeed, since its fundamental
representation is seven-dimensional, the only candidate is its
eight-dimensional spinor representation, but it is {\sl irreducible}
even in the massless case, thus it cannot split into two
four dimensional ones.
However, we note that (4.6) implies that these
$\o(5,2)$ generators act as matrix {\sl differential} operators on
$\psi$, i.e. the  representation given is {\sl not} a matrix
representation.\foot{
The same remark applies to the representation
obtained in Sec.~4  in the case of four-dimensional Minkowski
space:imposing the chiral projection on $\psi$, we
get a $2$-dimensional representation, though $\o(4,2)$ has no
$2$-dimensional matrix representation.
}
Put in a slightly different way, the solution space of the
massless Dirac operator in $5$ dimensions is $4\times$infinite
dimensional.

\goodbreak

%%%%%%%%%%%%%%%%%%%%%%%%%%%%%%%%%%%%%%%%%%%%%%%%%%%%%%
%%%%%%%%%%%%%%%%%%%%%%%%%%%%%%%%%%%%%%%%%%%%%%%%%%%%%%
\chapter{
Miscellany and discussion
}
%%%%%%%%%%%%%%%%%%%%%%%%%%%%%%%%%%%%%%%%%%%%%%%%%%%%%%
%%%%%%%%%%%%%%%%%%%%%%%%%%%%%%%%%%%%%%%%%%%%%%%%%%%%%%

As explained in Ref.~[6], a Maxwell term can be consistently
included into the FCI and Eq.~(1.9) becomes thus modified, as
$$
\sqrt{-g}\,
\epsilon_{\mu\nu\rho\sigma}\,\xi^\rho\nabla_\tau f^{\tau\sigma}
+\kappa f_{\mu\nu}
=e\sqrt{-g}\,
\epsilon_{\mu\nu\rho\sigma}\,\xi^\rho j^\sigma
\quad
\and
\quad
\partial\mathstrut_{[\mu}f\mathstrut_{\nu\rho]}=0
\equation
$$
(all indices running from $0$ to $3$).
This equation descends, as before, to the quotient, $Q$. In the
Minkowski case, for example, it reads
$$
\kappa B=-e\varrho,
\qquad
\partial_iB+\kappa E_i=e\,\epsilon_{ij}\,J^j
\qquad
\and
\qquad
\vnabla\times\vE+\partial_tB=0,
\equation
$$
generalizing the FCI (1.1) in the non-relativistic framework.
Clearly, the new system, Eqs~(10.2) and (1.4,5),
is still invariant with respect to the $\xi$-preserving conformal
transformations:
the RHS of Eq.~(10.1) is plainly invariant
under the rescalings
$g_{\mu\nu}\to\Omega^2g_{\mu\nu}$
and
$\xi^\mu\to\xi^\mu$
as a short calculation shows that the current scales as
$
j^\mu\to\Omega^{-4}j^\mu
$,
which is precisely compensated for by
$\sqrt{-g}\to\Omega^4\sqrt{-g}$.
Now, in $n$ dimensions, the Maxwell term scales as
$$
{\nabla}_\tau f^{\tau\sigma}\to
\Omega^{-4}{\nabla}_\tau f^{\tau\sigma}
+(n-4)\Omega^{-3}f^{\tau\sigma}\partial_\tau\Omega,
\equation
$$
with the sought conformal weight (the last term
vanishes here, as $n=3+1$).
Thus the LHS of Eq.~(10.1) is also invariant. So,
the combined LL-Maxwell-CS system is, indeed,
Schr\"odinger symmetric.

\goodbreak

It is worth mentioning that the non-relativistic conformal invariance
is maintained when modifying the massless Dirac equation (2.2) by an
`anomalous' term,
$$
\D\psi=ik\smallover\sqrt{-g}/2\,
\epsilon_{\mu\nu\rho\sigma}\gamma^\mu\xi^\nu f^{\rho\sigma}\psi,
\equation
$$
where $k=\const$, because both sides scale with the weight
$\Omega^{-5/2}$.
This amounts to modifying, in the previous developments, the covariant
derivative according to
$$
D_\mu
\to
D_\mu
-
ik\smallover\sqrt{-g}/2\,
\epsilon_{\mu\nu\rho\sigma}\xi^\nu f^{\rho\sigma},
\equation
$$
yielding a non-relativistic version of the model studied in Ref.~[21].

A peculiar feature of our non-relativistic `Kaluza-Klein' approach to
Chern-Simons theory is that it lacks an action principle for deriving the
four-dimensional form of the FCI, Eq.~(1.9). This difficulty is a
consequence of the {\sl null} character of the fibration over
Galilei spacetime; it forced us to deal essentially with the field
equations themselves in the study of the coupled LL-CS system.
It is worth noticing that, on the other hand, Carroll, Field and
Jackiw~[22] have discovered a four-dimensional version of Chern-Simons
theory admitting an action principle in the relativistic case, which is
governed by a space-like fibration discussed in Sec.~8 in connection with
relativistic spinor vortices.

Let us finally mention that our non-relativistic spinor vortices studied
in Secs~6 and~7 turn out to be different from those found before by
Leblanc et al.~[23], which are rather embedded scalar solutions.

\vfill\eject

%%%%%%%%%%%%%%%%%%%%%%%%%%%%%%%%%%%%%%%%%%%%%%%%%%%%%%
%%%%%%%%%%%%%%%%%%%%% Appendix %%%%%%%%%%%%%%%%%%%%%%%
%%%%%%%%%%%%%%%%%%%%%%%%%%%%%%%%%%%%%%%%%%%%%%%%%%%%%%

\centerline{\bf Appendix A}

\bigskip

Consider a $4$-dimensional spin manifold $(M,g)$ with
Clifford relations
$\gamma^{(\mu}\gamma^{\nu)}+g^{\mu\nu}=0$ and denote
$\nabla\psi$ the covariant
derivative of a spinor field $\psi$, Eq.~(2.1).
The spin-curvature, defined by
$$
\nabla_\mu\nabla_\nu\psi
-
\nabla_\nu\nabla_\mu\psi
=
\gamma(R)_{\mu\nu}\psi ,
\eqno(A.1)
$$
reads
$$
\gamma(R)_{\mu\nu}
\equiv
\4\gamma^\alpha\gamma^\beta R_{\alpha\beta\mu\nu}.
\eqno(A.2)
$$
(We use the convention
$
R_{\alpha\beta\mu}^{\;\cdot\;\cdot\;\cdot\lambda}
=
\partial_\alpha\Gamma^\lambda_{\beta\mu}
+ \cdots$.)
A tedious calculation (using the Bianchi identities
and the symmetries of the Riemann tensor) leads to
$
R_{\mu\nu\alpha\beta}\gamma^\nu\gamma^\alpha\gamma^\beta
=
-2R_{\mu\nu}\gamma^\nu$
where the Ricci tensor is defined by
$
R_{\beta\mu}=R_{\alpha\beta\mu\nu}\,g^{\alpha\nu}
$, hence, in view of Eq.~(A.2) to
the useful formula
$$
\gamma^\mu\gamma(R)_{\mu\nu}
=
\2\gamma^\mu R_{\mu\nu}.
\eqno(A.3)
$$
Recall the definition of the Lie derivative of a spinor
field, $\psi$, with
respect to a  vector field,~$X$, Eq.~(3.2),
$$
L_X\psi
\equiv
\nabla_X\psi -\2\gamma(d\barX)\psi
\eqno(A.4)
$$
where we have put $\barX\equiv{}g(X,\,\cdot\,)$ as a
shorthand notation
and used the convention (A.2).
{}From now on assume that $X$ be an infinitesimal conformal
transformation,
i.e.
$$
L_X g_{\mu\nu}=2\nabla_{(\mu}X_{\nu)}=k\,g_{\mu\nu}
\eqno(A.5)
$$
(with
$
k=\2\nabla_\mu X^\mu
$).
Calling $\nablaslash\equiv\gamma^\mu\nabla_\mu$ the Dirac
operator, we find,
with the help of Eqs~(A.4) and (A.1), that
$$
\displaylines{
[L_X,\nablaslash]\psi
=
X^\nu\nabla_\nu(\gamma^\mu\nabla_\mu\psi)
-
{1\over2}\gamma(d\barX)\gamma^\mu\nabla_\mu\psi
-\gamma^\mu\nabla_\mu(X^\nu\nabla_\nu\psi)
+
{1\over2}\gamma^\mu\nabla_\mu(\gamma(d\barX)\psi)
=\cr
X^\nu\gamma^\mu\gamma(R)_{\nu\mu}\psi
+
{1\over8}[\gamma^\mu,[\gamma^\alpha,\gamma^\beta]]
(\nabla_\alpha X_\beta)\nabla_\mu\psi
-
\gamma^\mu(\nabla_\mu X^\nu)\nabla_\nu\psi
+
{1\over8}\gamma^\mu[\gamma^\alpha,\gamma^\beta]
(\nabla_\mu\nabla_\alpha X_\beta)\psi.
\cr}
$$
We then compute separately each term in this last expression.
By using
$$
[\gamma^\mu,[\gamma^\alpha,\gamma^\beta]]
=
8\gamma^{[\alpha} g^{\beta]\mu}
$$
together with the partial result
$$
\gamma^\mu[\gamma^\alpha,\gamma^\beta]
\nabla_\mu\nabla_\alpha X_\beta
=
-2\nabla^\mu\nabla_\mu X_\nu\gamma^\nu
-
4\gamma(R)_{\mu\nu}\gamma^\mu X^\nu
+
4\gamma^\mu\partial_\mu k
$$
and (taking the divergence of both sides of Eq.~(A.5))
$$
\nabla^\nu\nabla_\nu X_\mu
+
R_{\mu\nu}X^\nu
=
-
\partial_\mu{k}
$$
one finds, using Eq.~(A.3), that
$
[L_X,\nablaslash]\psi
=
-{1\over2}R_{\mu\nu}\gamma^\mu X^\nu \psi
-{1\over2}\gamma^\nu(\nabla_\mu X_\nu+\nabla_\nu X_\mu)\nabla^\mu\psi
+{1\over4}(\partial_\mu{k}+ R_{\mu\nu}X^\nu)\gamma^\mu\psi
+{1\over4}R_{\mu\nu}\gamma^\mu X^\nu \psi
+{1\over2}\partial_\mu k\,\gamma^\mu\psi
$
and, finally,
$$
[L_X,\nablaslash]\psi
=
-\2 k\,\nablaslash\psi + \smallover 3/4 \gamma(dk)\psi
\eqno(A.7)
$$
where $\gamma(dk)\equiv\gamma^\mu\partial_\mu k$.
Notice the disappearance of the terms involving the curvature
in this last expression.

\vfill\eject

%%%%%%%%%%%%%%%%%%%%%%%%%%%%%%%%%%%%%%%%%%%%%%%%%%%%%%
%%%%%%%%%%%%%%%%%% the references %%%%%%%%%%%%%%%%%%%%
%%%%%%%%%%%%%%%%%%%%%%%%%%%%%%%%%%%%%%%%%%%%%%%%%%%%%%

\centerline{\bf\BBig References}

\bigskip

\reference
Z.~Zhang, T.~Hanson and S.~Kivelson,
Phys. Rev. Lett. {\bf 62}, 82  (1989); {\bf 62} 980 (E) (1989).

\reference
Y.-H.~Chen, F.~Wilczek, E.~Witten and B.~Halperin,
Int. Journ. Mod. Phys. {\bf B3}, 1001 (1989).

\reference
H.~B.~Nielsen and P.~Olesen,
Nucl. Phys. {\bf B61}, 45 (1973).

\reference
J.~Hong, Y.~Kim and P.~Y.~Pac,
Phys. Rev. Lett. {\bf 64}, 2230 (1990);
R.~Jackiw and E.~Weinberg,
Phys. Rev. Lett. {\bf 64}, 2234 (1990);
R.~Jackiw, K.~Lee and E.~Weinberg,
Phys. Rev. {\bf D42}, 2234 (1990).

\reference
R.~Jackiw and S.-Y.~Pi,
Phys. Rev. Lett. {\bf 64}, 2969 (1990);
Phys. Rev. {\bf 42}, 3500 (1990);
see Prog. Theor. Phys. Suppl. {\bf 107}, 1 (1992) for a review.

\reference
C.~Duval, P.~A.~Horv\'athy and L.~Palla,
Phys. Lett. {\bf B325}, 39 (1994).

\reference
C.~Duval, P.~A.~Horv\'athy and L.~Palla,
to appear in PRD.

\reference
J.-M.~L\'evy-Leblond,
Comm. Math. Phys. {\bf 6}, 286 (1967).

\reference
C.~Duval, G.~Burdet, H.-P.~K\"unzle and M.~Perrin,
Phys. Rev. {\bf D31}, 1841 (1985);
C.~Duval, G.~Gibbons and P.~Horv\'athy,
Phys. Rev. {\bf D43}, 3907 (1991), and references therein.

\reference
H.~P.~K\"unzle and C.~Duval,
Ann. Inst. H. Poincar\'e {\bf 41}, 363 (1984) ;
Class. Quant. Grav. {\bf 3}, 957 (1986).

\reference
C.~Duval,
in {\it Proc. XIVth Int. Conf. Diff. Geom. Meths. in Math. Phys.},
Salamanca '85, Garcia, P\'erez-Rend\'on (eds), Springer
Lecture Notes in
Math. {\bf 1251}, p.~205 Berlin (1987);
in {\it Proc. '85 Clausthal Conf.}, Barut and Doebner (eds),
Springer Lecture Notes in Physics {\bf 261}, p.~162  (1986).

\reference
C.~Duval, P.~A.~Horv\'athy and L.~Palla,
Phys. Rev. {\bf D50}, 6658 (1994).

\reference
Y.~M.~Cho, J.~W.~Kim, and D.~H.~Park,
Phys. Rev. {\bf D45}, 3802 (1992).

\reference
Y.~Kosmann,
Ann. di Mat. Appl. {\bf 91}, 317 (1972);
J.-P.~Bourguignon,
Rendiconti Sem. Mat. Univ. Politecn. Torino {\bf 44}, 3 (1986).

\reference
R.~Jackiw,
Phys. Today {\bf 25}, 23 (1972);
U.~Niederer,
Helv. Phys. Acta {\bf 45}, 802 (1972);
C.~R.~Hagen,
Phys. Rev. {\bf D5}, 377 (1972);
G.~Burdet and M.~Perrin,
Lett. Nuovo Cim. {\bf 4}, 651 (1972).

\reference
G.~Burdet, J.~Patera, M.~Perrin and P.~Winternitz,
J. Math. Phys. {\bf 19}, 1758 (1978).

\reference
J.-M.~Souriau,
Ann. Inst. H. Poincar\'e {\bf 20A}, 315 (1974).

\reference
C.~Duval,
Ann. Inst. H. Poincar\'e {\bf 25A}, 345 (1976).

\reference
Z.~F.~Ezawa, M.~Hotta and A.~Iwazaki,
Phys. Rev. Lett. {\bf 67}, 411 (1991);
Phys. Rev. {\bf D44}, 452 (1991);
R.~Jackiw and S.-Y.~Pi,
Phys. Rev. Lett. {\bf 67}, 415 (1991);
Phys. Rev. {\bf D44}, 2524 (1991).

\reference
H.~W.~Brinkmann,
Math. Ann. {\bf 94}, 119 (1925).

\reference
M.~Torres,
Phys. Rev. {\bf D46}, R2295 (1992).

\reference
S.~M.~Carroll, G.~B.~Field and R.~Jackiw,
Phys. Rev. {\bf 41}, 1231 (1990).

\reference
M.~Leblanc, G.~Lozano and H.~Min,
Ann. Phys. (N.Y.) {\bf 219}, 328 (1992).

\bye